\documentclass[11pt,a4paper]{article}
\usepackage{amssymb, amsmath, graphicx, subcaption, cite}
\usepackage{float}
\usepackage{graphpap,epsfig,textcomp}
\usepackage{epstopdf}
\usepackage{tabularx}
\usepackage[utf8]{inputenc}
\usepackage[english]{babel}
\usepackage{hyperref}
\usepackage[nottoc]{tocbibind}

\usepackage{multirow}
\usepackage{multicol}
\begin{document}
\begin{center}{\Large {\bf  Magnetic relaxation in the monolayer of ferromagnetic material}}\end{center}

\vskip 0.5cm

\begin{center} {\large {\it {Ishita Tikader and Muktish Acharyya }}
\vskip 0.1cm
{\large {Department of Physics, Presidency University, Kolkata, India}}}
\end{center}

\vskip 0.2cm


\tableofcontents	

\newpage


\noindent {\bf Abstract:} 
This article present a systematic review of studies focused on relaxation behaviours of the magnetic systems. The theoretical as well as experimental investigations in this specific issue are briefly discussed here. A major part of this article focuses on recent stochastic (Monte Carlo) simulation-based studies into the impact of dynamics, applied boundary conditions and the  structure of the system on the decay of magnetisation of ferromagnetic monolayers, using the model of widely used two-dimensional Ising ferromagnet. Our discussions include two types of geometrical deformations of Ising ferromagnetic systems, namely `area-conserving' and `area-non-conserving'. The simulation is conducted for the systems under open boundary conditions (OBC) and periodic (PBC) boundary conditions. A comperative study between Metropolis  and Glauber dynamical algorithms is also conducted through out our obsevation. 
 The decay of instantaneous magnetisation indicated the exponential nature  of magnetic relaxation and demonstrates the remarkably significant dependence of the time scale (namely, the relaxation time $\tau$) upon the aspect ratio ($R =\frac {\rm Length }{\rm Breadth}$). The  power-law type of relationship between $\tau$ and $R$ ($\tau \sim R^{-s}$) can be established in the large aspect ratio ($R$) limit. Further analysis expose the linear decrement in the exponent associated with power-law relation ($s$) due to a gradual increase of temperature of the system ($s = aT + b$) near critical temperature. In addition, the temporal growth of the spin-flip density is critically studied for surface and core region of the lattice. The saturated value of spin-flip density shows a strong logarithmic  variation $f_{sd} = a + b~\log(L)$  with size ($L$) of the lattice for core region as well as surface. Our study may prompt the experimentalists for conducting experimental exploration  leading the adavancement of \emph{magnetic coating} on the magnetic stripe cards to achieve the faster response time.



\vskip 2cm

\noindent {\large \bf Objectives:}\\

\begin{itemize}
    \item The magnetic relaxation and the theoretical formulation.
    \item Relaxation of magnetic monolayers studied through simulation.
    \item Role of the lattice geometry on the relaxation.
    \item How do the boundary conditions and dynamical rules affects relaxation?
    \item Connection of the density of spin flips and relaxation.
    \item Experimental scenario of magnetic relaxation.
    \item Magnetic coating and its technological importance.
    
\end{itemize}


\vskip 2 cm


\newpage


\section {Preface}


Relaxation is a ubiquitous phenomenon that occurred in a wide variety of real life systems.
Particularly, depending on the temperature, the interacting many-body systems (in equilibrium with
thermal resevoir) show intriguing relaxation behaviours.
The cooperatively interacting thermodynamic systems, has drawn the attention of researchers due to the relaxational nature as they strive to reach an equilibrium or steady state. When subjected to an external perturbation such as externally applied magnetic field, electric field or stress etc., the system requires a certain duration of time to recover its original and well described stationary state, after the withdrawal of external disturbance. The investigation on ferromagnetic relaxation has emerged as captivating and well-establish field of research, attracting both theoretical and experimental investigations over past few decades. 

When ferromagnetic samples is disturbed by a magnetic perturbation, it stats to relax \cite{article2} and gradually returns back towards its original thermodynamic equilibrium state, when the of external magnetic perturbation is eliminated. Some experimental findings on the magnetic relaxation of various samples are briefly reviewed here. In an experiment by Godfrin et al.(1980), the relaxation process of surface magnetisation was studied in $He^3$ by using Nuclear Magnetic resonance technique within a confined geometry \cite {godfrin:1980}. An experimental exploration of the magnetic relaxation in superconducting ${\rm YBa_2 Cu_3 O_7}$ sample was conducted in the temperature range 77-95K \cite{zazo1993experimental}. The relaxation of magnetic moments was experimentally investigated in the CNT-enhanced composite of Fe-nanoparticles \cite{Rosa:2019}. 
 The dysprosocenium complex based rare-earth SIMs \cite {cheisa:2020} as well as single molecular magnets (SMMs) \cite{PhysRevLett.125.117203} exhibit slow magnetisation decay or magnetic relaxation. The experimental evidences of non-Arrhenius (significanly deviating from Neel-Arrhenius law) magnetic relaxation in highly homogeneous amorphous thin film of ${\rm CoZrDy}$  are also reported \cite{Suran:1998}. Both longitudinal and transverse rates of ferromagnetic relaxation in a micron-sized sample have been measured by utilizing ferromagnetic resonance force microscopy \cite{PhysRevB.67.220407}.

Now a partinent question to be addressed here how one can investigate the relaxation in a simple ferromagnetic monolayer
or two dimensional (2D) magnets? The ferromagnetic monolayer can be a realization of a 2D ferromagnetic system. So researchers may employ the spin-1/2 (in the limit of very large anisotropy) 
Ising ferromagnet as a prototypical model. The relaxation of total z component of magnetisation $M_z$ in near-Ising ferromagnet namely dysprosium ethyl sulphate or DyES was extensively studied and analysed in the past \cite{PhysRev.187.690}. The classical Ising model does not possess an intrinsic dynamics. However, the dynamical equations using simplest approximation corresponding to mean-field treatment, are explicitly formulated \cite{suzuki:1968} for Ising model in two and three dimensions, specially treated by Glauber protocol \cite{Glauber:2004}. This averaged or mean-field type of dynamical equation has been used as the simplified model to explore various phenomena related to the dynamic behaviour such as relaxational behaviour of the spin-spin correlation of the relaxation of magnetisation.

 Several noteworthy theoretical and simulation-based studies focusing on magnetic relaxation are briefly discussed here, so that one can understand the importance of Ising model to explore relaxation behavior. How the correlation fuctions of the spins in 1D Ising model decay with time paticularly in low temperature limit has been investigated using Glauber kinetic algorithm \cite{PhysRevE.53.458}. Extremely long Monte Carlo simulation study of dynamics around critcal region have been performed \cite{PhysRevE.93.022113} using FPGA-based computing devices to explore the linear relaxation process in a large (up to $2048\times 2048$) 2D Ising ferromagnetic model and determine the dynamical exponent (critical) $z$. The relaxation behavior of the internal energy in 2D Ising system having lattice size $L=5120$ at temperature region above $T_c$ was studied by using supercomputer-conducted MC simulational dynamics and calculated the critical exponents related to the corresponding linear and nonlinear relaxation times \cite{PhysRevB.47.11499}. The relaxation time corresponds to simple linear response and nonlinear response of order parameter in two and three-dimensional dynamical Ising model are critically analysed \cite{PhysRevB.35.394} by using Pad\'e approximation. Muller-Krumbhaar and Landau have investigated fascinating tricritical relaxation behavior in Glauber kinetic Ising model in three-dimension \cite{PhysRevB.14.2014} and observed the exponential nature of relaxation below the tricritical temperature $ T_t $ as well as the slowing down of relaxation time (CSD) at tricritical point. The relaxation behavior of order parameter is critically examined in the Ising model on a small-world  network \cite{PhysRevE.71.036103} by using numerical simulations and very slow relaxation process, caused by strong long-range type interactions (where $\frac {J_2} {J_1} \gg 1$) is clearly identified. Tomita and Nonomura (2018) have examined the nonequilibrium relaxation (NER) behavior at critical point $T_c$ in Ising system by using cluster-flip dynamics. They have reported \cite{PhysRevE.98.052110} the stretched-exponential nature of relaxtion for lower dimension $ d\leq 3 $. 

Tomita and Miyashita (1992) \cite{PhysRevB.46.8886} have extensively investigated the magnetic field and system size dependence of the relaxation time for the metastably ordered state of the 2D Ising model while encountering the longitudinally unfavorable field by the help of Ising Machine m-TIS2. The extremely slow decay of average magnetisation (relaxation) described by an inverse logarithmic function has been demonstrated \cite {PhysRevE.66.056114} in a specially designed Ising spin chain by implementing a simplified or toy model. The nonequilibrium relaxation of magnetisation and energy at critical point in a two-dimensional Ising model were analysed \cite{PhysRevE.57.6548} by series expansion method and MC simulational scheme which led to estimate the dynamical critical exponent $z$.

The well known cluster theory is developed to study the dynamical properties near critical point of kinetic Ising model \cite{PhysRevB.12.5261} obeying Glauber protocol. The presence of disorder or the quenched  impurities of nonmagnetic material  plays an crucial role in the decay of the magnetisation in the ferromagnetic system. Focusing on this point, Oerding systematically investigated \cite{1995JSP:78:Oerding} the impact of randomly quenched nonmagnetic impurities on the relaxation phenomenon exhibited by the Ising ferromagnets and observed that the presence of impurity can produce the continuous spectra of relaxation time while the relaxtion of a system without any impurity can be charaterized by a discrete set. A scientifically relevant question should be raised in the context of statistical mechanics. What type of statistical distribution function is expected to come out when magnetic relaxation times are  measured for uncorrelated samples? M\'elin recorded the data of time-scales related to the reversal of magnetization in magnetic grains and noted the Gaussian nature of the histogram of $P(\ln {\tau})$ the logarithmic value of the relaxation time \cite{melin1996magnetic}. Relaxation has also been studied by monitoring the decay of ``damage"  or ``Hamming distance" in stochastic dynamics (MC) with the help of the spreading of damage technique. Grassberger and Stauffer (1996) predicted that decay of excess magnetisation in 3D and 4D Ising ferromagnet is simply exponential by nature, observed in ferromagnetic as well as paramagnetic regime \cite {GRASSBERGER1996171}. But in the two dimensional system, the decay of `Damage' can be characterized by a stretched-exponential function only below $T_c$. The research, related to the relaxation kinetics in the Ising like ferromagnetic model is not only bounded by spin-1/2 system. It would be extremely fascinating to investigate the relaxation process in the models designed by the  higher values of the spin. The dynamic properties of magnatisation in spin-1 Ising ferromagnet (known as isotropic BEG model) around the critical point during 2nd-order or continuous phase transition is also examined \cite {ERDEM20082273} and magnetic dispersion along with absorption factor is determined as well. The antiferromagnetic system also exhibits relaxation behavior. Nowak and Usadel investigated slow relaxtion of remanent magnetisation \cite {Nowak89} in diluted Ising-type antiferromagnets by Monte Carlo simulation and found that in low temperature the relaxation process occurs in domain walls.
Recently, the nonequilibrium evolution near the pahse boundary of three
dimensional Ising ferromagnet and the relaxation behaviour near the first
order phase transition line has been studied \cite{Li:2023}.

With this coincise overview on the study of relaxation mechanism in the kinetic Ising system, we present several unique and fascinating outcomes of our recent investigations on the dynamical evolution of magnetisation in the kinetic Ising monolayer (i.e., 2D system) specially focusing on the impact of various geometrical structures. We also highlight some results on the  relaxation behaviour observed in Monte Carlo simulation governed by different simulational algorithms. The effects of the different applied boundary conditions  are also reviewed in this  article.



\section {Decay of magnetisation in Ising ferromagnet: Mean field formalism}

Historically, the simplest model of the kinetic Ising ferromagnet to study the dynamical responses, has been developed
in 1968 \cite{suzuki:1968}. The approach was primarily of mean field type imposing Glauber dynamical rules in Ising
ferromagnet. Let us try to give a brief idea about such kind of dynamical equation.

The model of ferromagnetically interacting Ising spins (spin-$\frac{1}{2}$) is considered where spin variables are denoted by $\{ \sigma_i \}$. Individually, each spin is allowed to have only the two discrete values, e.g., +1 and -1. The system exchanges energy with a large heat reservoir maintained at a constant value of the temperature $T$ which leads to spontaneous spin-flips. Here $\Omega_i(\sigma_i)$ represents the transition probability of spin state of $i^\text{th}$ spin. This is not governed by $\sigma_i$ only, but also affected by the interacting neighbouring spins. The master equation that describes the dynamic evolution of probability, can be formulated for the Ising model and expressed as follows 
\begin{equation}\label{eqn:1}
    \frac{dP_r(\sigma_1,...\sigma_i,...\sigma_N;t)}{dt}=- \sum_{i} \Omega_i(\sigma_i)P_r(\sigma_1,...\sigma_i,...\sigma_N; t) + \sum_{i} \Omega_i(-\sigma_i)P_r(\sigma_1,...- \sigma_i,...\sigma_N; t)
\end{equation}
where $P_r(\sigma_1,\sigma_2,....\sigma_N;t)$ denotes the chance of getting the spins in the particular arrangement $(\sigma_1, \sigma_2,.....\sigma_N)$. The very $\rm 1^{st}$ term in the RHS of the dynamical equation-\ref{eqn:1} arises due to the reduction in probability caused by reverting out of spin and the second term corresponds to the increase in probability resulting from spin-flips occurring in the opposite direction. We know that the thermal equilibrium assures $\frac{dP_r}{dt} = 0$. According to the widely applied celebrated Principle of detailed balance, we may write the equality equation,
\begin{equation}\label{eqn:2}
   \Omega_i(\sigma_i)P_{\rm equil}(\sigma_1,...,\sigma_i,...\sigma_N) = \Omega_i(-\sigma_i)P_{\rm equil}(\sigma_1,...,- \sigma_i,...\sigma_N)
\end{equation}
where $P_{\rm equil}(\sigma_1...\sigma_N)$ represents the probability in the equilibrium state of the system. 
So, the Equation-\ref{eqn:2} can be rewritten as follows,
\begin{equation}\label{eqn:3}
    \frac{\Omega_i(\sigma_i)}{\Omega_i(-\sigma_i)} = \frac{P_{\rm equil}(\sigma_1,...-\sigma_i...\sigma_N)}{P_{\rm equil}(\sigma_1,...\sigma_i...\sigma_N)} 
\end{equation}
 The energy of the system with Ising spins (when subjected to a magnetic field, $B$) is represented by the following kind of Hamiltonian,
\begin{equation}\label{eqn:4}
	H_{\rm Ising}=-\sum_{<i,j>} J_{ij}\: \sigma_i \sigma_j - \sum_{i} B_{i}\: \sigma_i 
\end{equation}
Here the first term corresponds to the interaction energy between the nearest neighbouring pair $<ij>$ of spins and the second sum represents the energy caused by the externally applied magnetic field $B_i$ (in general function of the position). 

The chance $P_{\rm equil}$ of getting a system in an arrangement of the configuration `$C$' is governed by Gibbs distribution formula, 
\begin{equation}\label{eqn:5}
    P_{\rm equil}(C)= \frac{1}{Z} \exp{(-\beta E_{c})}
\end{equation}
where $\beta = 1/{k_BT}$; $k_B T$ denotes the product of Boltzmann constant $k_B$ and the system's temperature $T$ . $E_{C}                                                               $ is the energy of the configuration $C$. $Z$ defines the partition function.\\ 
Detailed balanced condition, (in equilibrium) metioned in the Equation-\ref{eqn:2} confirms the following mathematical expression,
\begin{equation}\label{eqn:6}
    \frac{\Omega_i(\sigma_i)}{\Omega_i(-\sigma_i)} = \frac{\exp{(-\beta \epsilon_i \sigma_i)}}{\exp{(\beta \epsilon_i \sigma_i)}}
\end{equation}    
where $\epsilon_i$, the local energy (cooperative and field) at a lattice site, is denoted by,
\begin{equation}\label{eqn:7}
	\epsilon_i= \sum_{j} J_{ij}\: \sigma_j + B_{i} 
\end{equation}
Considering that Ising spins are restricted to two discrete values (+1 or -1), we can express the aforementioned exponential terms in the form of hyperbolic functions using the properties of Pauli spin matrices.
\begin{equation}\label{eqn:8}
    \exp{(\pm \beta \epsilon_i\sigma_i)}= \cosh(\beta \epsilon_i) \pm \sigma_i\: \sinh(\beta \epsilon_i)= \cosh(\beta \epsilon_i)[1 \pm \sigma_i\: \tanh(\beta \epsilon_i)]
\end{equation}
Thus, we can rewrite the Equation-(\ref{eqn:6}) as,
\begin{equation}\label{eqn:9}
    \frac{\Omega_i(\sigma_i)}{\Omega_i(-\sigma_i)} = \frac{1-\sigma_i\: \tanh(\beta \epsilon_i)}{1+\sigma_i\: \tanh(\beta \epsilon_i)}
\end{equation}  
The transition probability can be mathematically expressed as follows using the Glauber's proposal \cite{Glauber:2004} 
\begin{equation}\label{eqn:10}
    \Omega_i(\sigma_i)=\frac{1}{2\tau_m} (1-\sigma_i\: \tanh(\beta \epsilon_i))
\end{equation}

Here the constant $\tau_m$, termed as the intrinsic microscopic relaxation time which depends on the temperature as well as the interacting spins in the nearest neighourhood.\\
By the defination of expectation value we may write (for $i^{th}$ spin), 
\begin{equation}\label{eqn:11}
    \langle {\sigma_i} \rangle = \sum_{\{ \sigma_i \}} \sigma_i\: P_r(\sigma_1,...,\sigma_i,..\sigma_N;t)
\end{equation}
The total time derivative of expectation value of $i^{\rm th}$ spin $\langle{\sigma_i} \rangle$ leads us as follows,
\begin{equation}\label{eqn:12}
    \frac{d\langle {\sigma_i} \rangle}{dt}= \sum_{\{ \sigma_i \}} \sigma_i\: \frac { dP_r(\sigma_1,...,\sigma_i,..\sigma_N;t)}{dt}
\end{equation}
Utilizing the master equation mentioned in the Equation-\ref{eqn:1} we can successively derive the following kinetic equation:
\begin{equation}\label{eqn:13}
    \tau_m \frac{d}{dt}\langle \sigma{_{i}}\rangle = -\langle \sigma{_{i}}\rangle + \langle \tanh{\beta \epsilon{_{i}}}\rangle
\end{equation}
This equation can be readily solved using the mean-field theory, which provides a basic yet effective approximation. In this approximation, the Equation-\ref{eqn:13} can be substituted with the following expression,
\begin{equation}\label{eqn:14}
    \tau_m \frac{d}{dt}\langle \sigma{_{i}}\rangle = -\langle \sigma{_{i}}\rangle +  \tanh{(\beta \langle {\epsilon_i}\rangle)}
\end{equation}
In the specific case of a spatially uniform applied magnetic field, we may write for $\langle {\epsilon_i} \rangle$ 
\begin{equation}\label{eqn:15}
    \langle {\epsilon_i} \rangle = \sum_{j} J_{ij}\: \langle {\sigma_j} \rangle + B 
\end{equation}
 In the mean-field approximation the kinetic equation(\ref{eqn:14}) would be reformulated for instantaneous average magnetisation m(t), where $\langle{\sigma_i} \rangle $ and $\langle{\sigma_j} \rangle$ are replaced by $m$ (Since the expectation value $\langle {\sigma_i} \rangle$ of spin at given site $i$ is independent of site in mean field approximation, so  $\langle {\sigma_i} \rangle = \langle {\sigma_j} \rangle = m$); 
\begin{equation}\label{eqn:16}
 \tau_m \frac{dm}{dt}= -m + \tanh{\beta (B+\lambda m)};\hspace{0.5cm} \hspace{0.1 cm}
 {\rm where,} \hspace{0.2cm}  \lambda=\sum_{j} J{_{ij}}
\end{equation}
In the context of zero magnetic field (with $B=0$), the equation(\ref{eqn:16}) can be linearized near the critical temperature ($T \rightarrow T_c$) (neglecting the higher order terms of `$\tanh$' function),
\begin{equation}\label{eqn:17}
\begin{split}
\tau_m \frac{dm}{dt} =-\zeta m;\\
{\rm where,} \hspace{0.2cm} 
\zeta = 1- \beta \lambda = 1-\frac{T{_{c}}}{T}; \\
T{_{c}}= {\dfrac {\sum J{_{ij}}}{k_B}}, {\rm ~ Critical\hspace{0.2cm}Temperature}
\end{split}
\end{equation}

 The explicit solution of the differential equation-(\ref{eqn:17}) is expressed as follows,  

\begin{equation}\label{eqn:18}
    m(t) \thickapprox \exp{(-\frac{t}{\tau_r})}
\end{equation}
It is evident that the relaxation behavior of magnetisation at a constant temperature can be effectively described by exponential decay.\\
Where,
\begin{equation}\label{eqn:19}
  \tau {_{r}}= \dfrac{\tau_m}{\zeta}=\dfrac{\tau_m T}{T-T{_{c}}}
\end{equation}
Here $\tau_r$ is signified as the macroscopic relaxation time arises due to the interaction of the spins under mean-field treatment. From the expression of $\tau_r$, we may interpret that $\tau {_{r}}\sim(T-T{_{c}})^{-1}$.  The relaxation time diverges as temperature approches the critical point ($T \rightarrow T_c$). This phenomenon is defined as \emph{`Critical slowing down (CSD)'}.
 
In the mean-field approximation the thermodynamic fluctuations are disregarded. However we should keep in mind that thermodynamic fluctuations play a crucial role in the transition from ordered ferromagnet to the disordered paramagnet. Therefore to investigate the relaxation behaviors while considering the influence of thermodynamical fluctuations, we need to employ a basic lattice based model and the MC scheme of simulation with specified governing dynamical rules.



\section {Model of Ising ferromagnet and applied Monte Carlo simulational methodology}
\label{model}

In this particular chapter, our analysis is primarily focused on the simulational studies based on stochastic dynamics \cite{Mllerkrumbhaar1973DynamicPO} of relaxation mechanism observed in the ferromagnetic materials. We are interested in the special kind of spin model, more precisely the Ising ferromagnetic model. The classical Ising model is a widely studied prototypical spin model for studying the equilibrium and nonequilibrium behaviors of the magnetic systems.

The magnetic monolayer (when no external field is applied to the system) can be described by the Hamiltonian of Ising model, expressed as
\begin{equation}
	 H_{\rm Ising}=-J\:\sum_{<ij>} \sigma_i \sigma_j 
	\label{eqn:Ising}
\end{equation}	
where discrete spin variable at each lattice site $\sigma_i$ ($ \forall ~ i $) can assume either value +1 (spin up) or -1 (spin down). The nearest-neighbour spins ($<ij>$) are uniformly coupled with interaction coefficient $J$. The value of $J$ has to be considered positive ($J>0$) to have truly ferromagnetic ground state of the aforementoned Hamiltonian.

Here, our system is a rectangular lattice of size ($L_x \times L_y$) and the aspect ratio is defined as $R=\frac{L_x}{L_y}$. The initial configuration is chosen as the ground state of ferromagnet (Ising) where the all spins have assumed up directions. This is analogous to the very high field (applied externally) configuration. However, at any finite temperature $T$, this should not be the equilibrium configuration of the system. The system must come down to equilibrium state just by relaxation.
The relaxation is studied by the random updating of the spins with Matropolis and Glauber protocol. The flipping of the spins are probabilistically considered.

The Metropolis probability \cite{Metropolis:1953} of the spin-flip  is given by, 
\begin{equation}\label{eqn:20}
    P(\sigma_i \rightarrow  -\sigma_i) = Min[1, \exp (-\frac{\Delta E}{k_BT})]
\end{equation}

Whereas, the Glauber rate for the spin filp is represented by,  
\begin{equation}\label{eqn:21}
    P(\sigma_i \rightarrow  -\sigma_i) = \frac{\exp (-\frac{\Delta E}{k_BT})}{1+ \exp (-\frac{\Delta E}{k_BT})}
\end{equation}
 Where the temperature of the system is symbolized by $T$ in $\rm J/k_B$ unit. $\Delta E$ denotes the difference in enegy between final and initial state caused by trial spin-flip and the energy is expressed in unit of $J$. The Boltzmann constant is denoted by $k_B$. For simplification, $k_B ~\rm and~ J$ are assumed be unit valued throughout our numerical study.
  
 In the updating of the spin, any of the above-mentioned methods can be executed as follows: the probability whether a spin will flip or not will be compared to a random number, having uniform distribution in the range of (0, 1). If the randomly generated number does not exceed the probability, only then the spin will flip.

In this way, total $N=L_x\times L_y$ numbers of randomly selected lattice sites (random updating scheme) will be updated by executing the above mentioned spin-flip protocols. Such random updates will constitute  MC Step per Site or MCSS.  
The time dependent magnetisation is formulated as:
\begin{equation}\label{eqn:22}
   \widetilde{m}(t)= \dfrac{\sum_{i=1}^{N} \sigma_{_{i}}}{N}. 
\end{equation}
The ensemble average of magnetisation at a time instant $m(t)$ is computed over a large number of uncorrelated samples.


\section {Numerical results}
\subsection{\bf Influence of geometrical structure, applied boundary conditions and dynamical algorithms on the relaxation of magnetisation}

In the begining, a square lattice of size $L{_{x}}=L{_{y}}$ is opted for observation, whose spin variable at each lattice site is initially aligned upwards i.e., $\sigma{_{i}}=+1;~\forall i$. One can imagine the scenario as a system is perturbed to the state of the saturation magnetisation by an externally applied strong magnetic field. After abrupt removal of the magnetic field the system returns back towards its equilibrium state at a temperature above transition point ($T>T{_{c}}$). That means the magnetisation of the system starts to decay with time and eventually achieves the state of zero magnetisation. In this present study the magnetic relaxation in 2D Ising ferromagnet is critically examined to explore the impact of the geometrical structure  and boundary condition of lattice on relaxation behavior. Moreover The roles of different dynamical protocols on relaxation are thoroughly investigated.      
  
\begin{figure}[h!]
\begin{center}
\includegraphics[angle=0,scale=0.63]{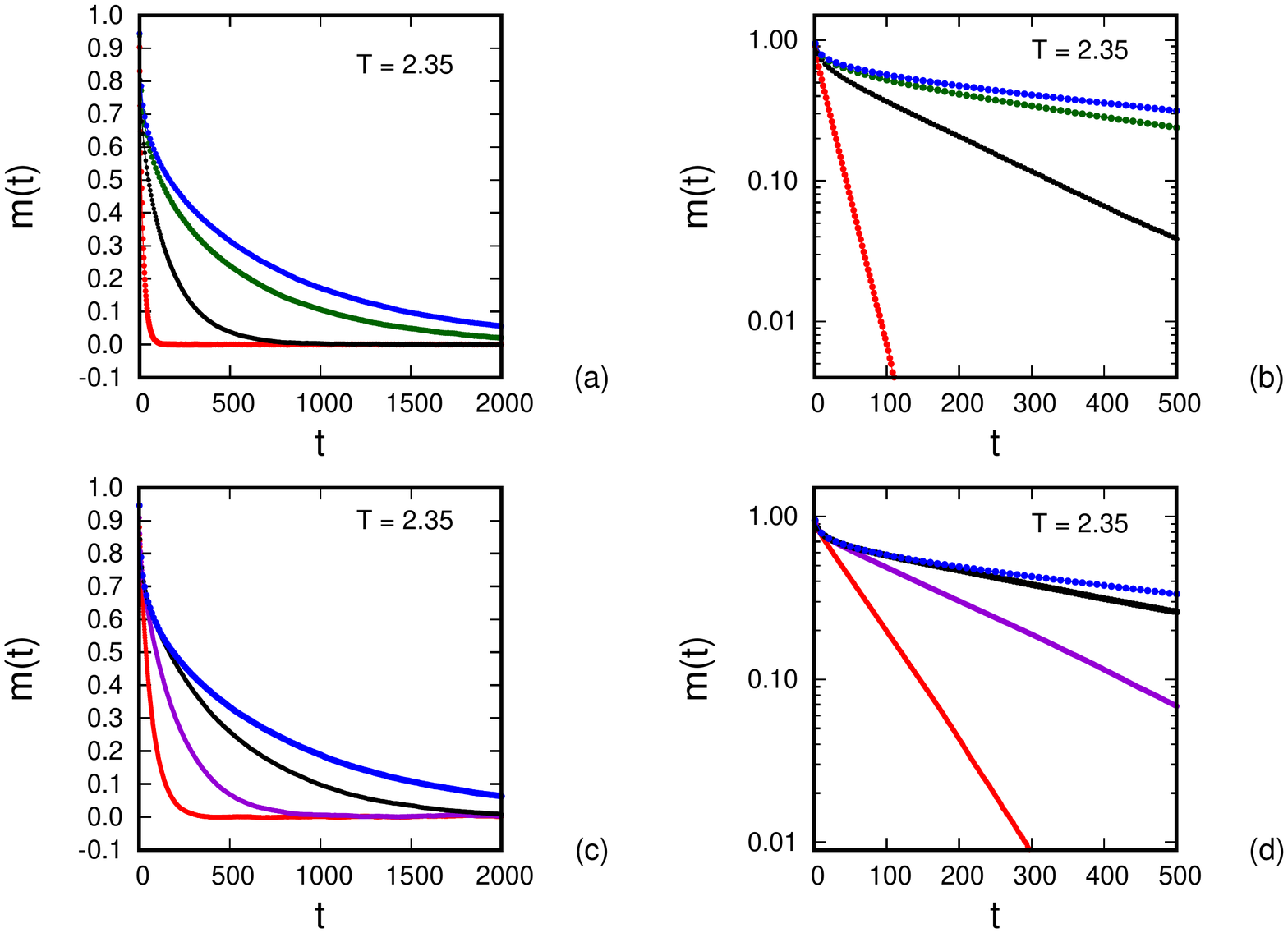}
\caption{Time evolution of magnetisation $m(t)$ at fixed temperature $T = 2.35~ \rm J/k_B$ for various system sizes. $m(t)$ obtained using Open Boundary Condition plotted in (a) the linear space and (b) the semi-log scale. 
Data obtained using Periodic  Boundary Condition also plotted in (c) the linear space and (d) semi-log scale. Collected from \cite{tikader2022effects}.}
\label{m-t-metro}
\end{center}
\end{figure}
The temporal decay of magnetisation is studied at a fixed temperature $T=2.35~\rm J/k_B$ above critical point and obtained results using Metropolis dynamical rule for different lattice sizes (by changing $L_y$ keeping value of $L_x$ fixed) are graphically demonstrated in the Figure-\ref{m-t-metro}. The Figure-\ref{m-t-metro}(a) depicts the time evolution of magnetisation for the systems with open boundary condition (OBC). The log value of magnetisation ($\log m$) shows a linear variation with time (Figure-\ref{m-t-metro}(b)), which implies the exponential decay of the $m(t)$. So, we may report $m(t) \sim \exp (-\frac {t}{\tau})$. The characteristic time regarding magnetic relaxation ($\tau$) termed as relaxation time of the system is determined by taking inverse of the slope of the fitted line of $\log m$ vs. time data points using linear regression method. Similarly, we have studied the relaxation (decay) of magnetisation for the Ising systems with periodic boundary condition (PBC) and reprensented in the Figure-\ref{m-t-metro}(c) (for linear-scale) and the Figure-\ref{m-t-metro}(d) (for log-scale). The relaxation time $\tau$ is also measured for the system with PBC in similar fashion (linear fit of $\log (m)-t$). Qualitatively similar outcomes are found for both OBC and PBC. It is remarkable that comperatively faster relaxation process is observed (analyzing the graphical representations in the Figure-\ref{m-t-metro}(b) and the Figure-\ref{m-t-metro}(d)) for the system with OBC, while keeping all other parameters (dynamics, temperature and system size) unchanged. 
\begin{figure}[h!]
\begin{center}
\includegraphics[angle=0,scale=0.63]{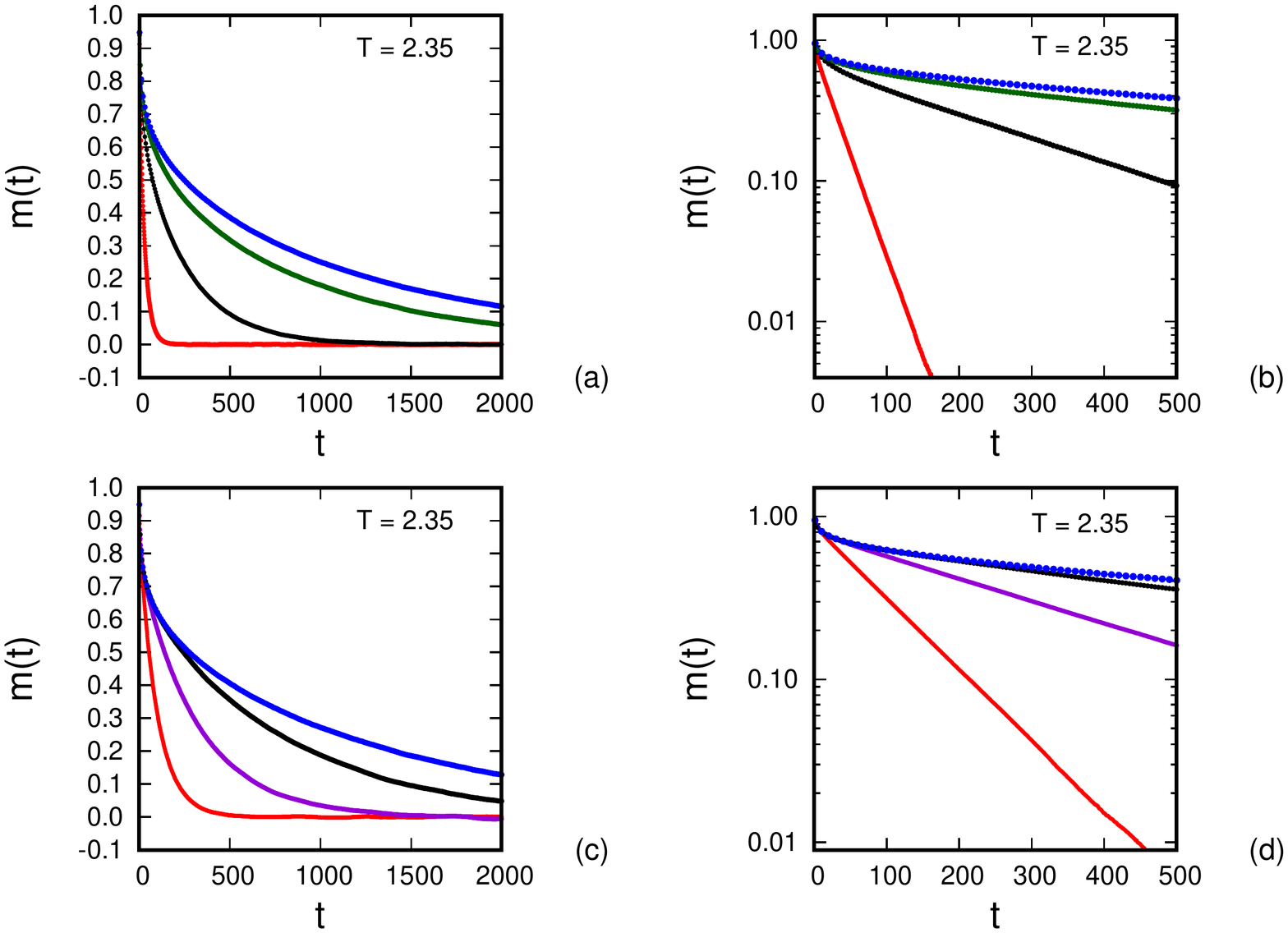}
\caption{ The evolution of magnetisation $m(t)$ with time at fixed temperature $T = 2.35 ~ \rm J/k_B$ for various system sizes. 
$m(t)$ obtained using Open Boundary Condition shown in (a) the linear plot and (b) the semi-log plot. 
$m(t)$ obtained using Periodic  Boundary Condition also presented in (c) the linear plot and (d) the semi-log plot. Collected from \cite{tikader2022effects}.}
\label{m-t-logm-t-glau}
\end{center}
\end{figure}

The investigation on magnetic relaxation is also carried out for the systems under OBC and PBC by implementing the Glauber protocol. Our observations are graphically represented in the Figure-\ref{m-t-logm-t-glau}. The temporal variation of magnetisation in linear scale and semi-log scale are depicted in the Figure-\ref{m-t-logm-t-glau}(a) and the Figure-\ref{m-t-logm-t-glau}(b) respectively for the systems under open boundary condition(OBC). Similarly, the Figure-\ref{m-t-logm-t-glau}(c) and the Figure-\ref{m-t-logm-t-glau}(d) provide the graphical representation of data obtained by employing periodic boundary condition (PBC) in linear-scale and semi-log scale respectively. Here also, the systems with open boundary condition exhibit the faster relaxation process. In terms of qualitative analysis, it must be noted that while applying OBC, the spins located at the boundary experience the interactions with only three spins at nearest-neighbour site and those on the corner are surrounded by two interacting nearest-neighbours in a two-dimensional lattice. Whereas in the 2D lattice with PBC, the spin at each lattice site have four interacting nearest-neighbours. Consequently, the system under OBC is more susceptible to spin reversal, thereby facilitating  faster relaxation process compared to PBC.

 Moreover, it is obvious from the results that the system, when utilizing Metropolis dynamical rules, achieves the thermodynamic equilibrium at faster rate in  comparison to the system employing Glauber protocol. The underlying reasons for this discrepancy is readily apparent. The Metropolis dynamical rule incorporates a probability mechanism that enables spin-flips with unit probability when spin-flip causes the negative change in energy. In contrast, such provision is absent in the Glauber protocol, resulting relatively slower relaxation process.
 
An itriguing question should be addressed in this context. How different will be magnetic relaxtion if we deform the geometrical shape of the 2D Ising system, precisely when a square lattice is transfromed into rectangular structure. The change in relaxation time due to geometrical deformation is duly noted for (i) non-conserved area (reducing $L_y$ for fixed $L_x$) and (ii) conserved area  (keeping $L_x \times L_y$ constant).
\vskip 0.2cm
\subsubsection{\bf{Rectangular geometry: Non-conserved area}} 
\begin{figure}[h!]
\begin{center}

\includegraphics[angle=0,scale=0.63]{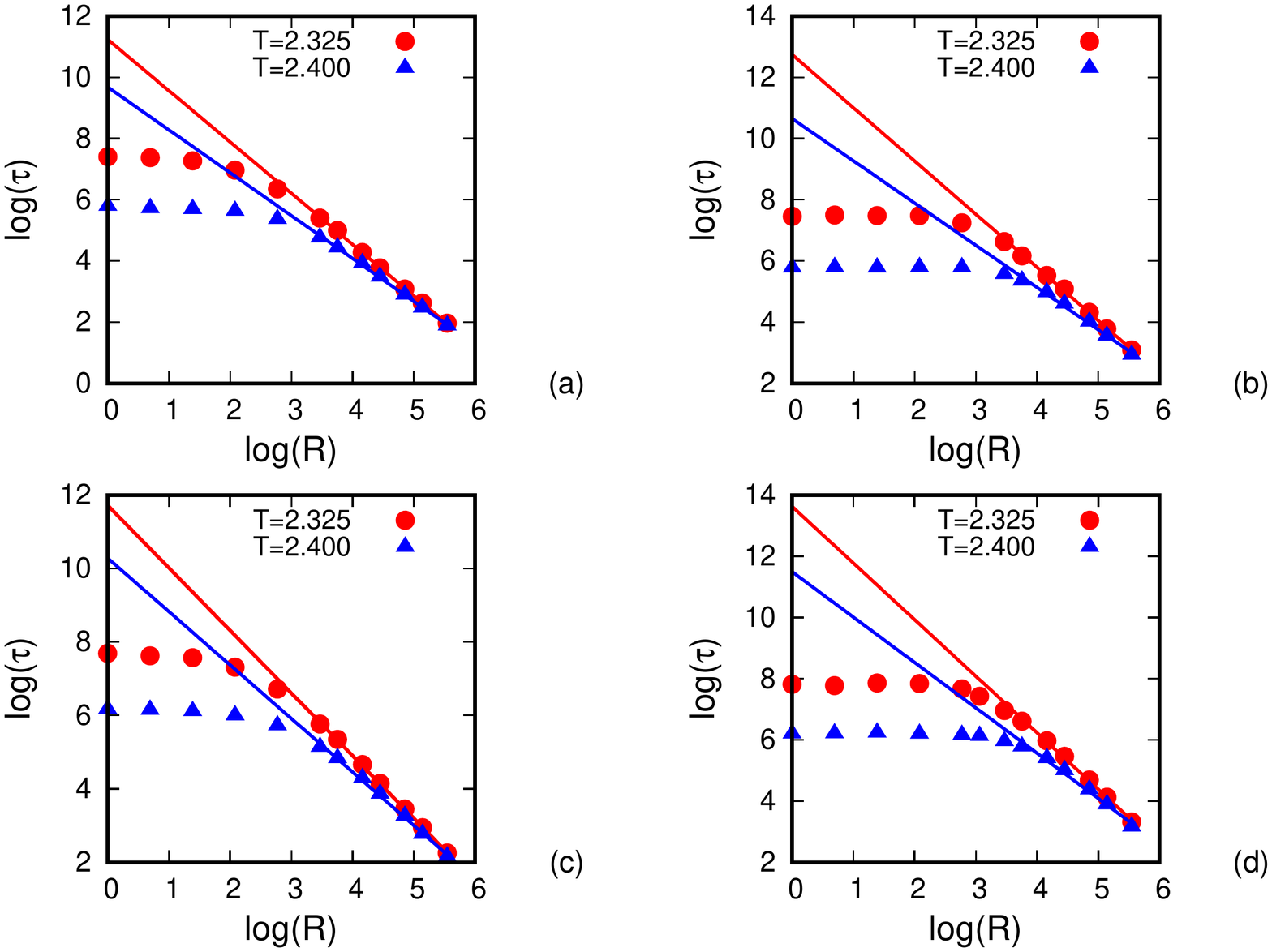}

\caption{The geometrical dependence of relaxation time. Variation of $\tau$ with the aspect ratio ($R$) depicted in a double-logarithmic plot. Red circles symbolize the data obtained at $T=2.325~J/k_B$ and Blue triangles indicate the temperature $T=2.4~J/k_B$ in the context of `area non-conserving deformation'. Linear fit ($\log(\tau)=-s\times \log(R)+r$) is indicated by Red line for $T=2.325~J/k_B$ and Blue line for $T=2.4~J/k_B$.  Results obtained using
(a) OBC and Metropolis dynamics, 
(b) PBC and Metropolis dynamics,
(c) OBC and Glauber protocol and
(d) PBC and Glauber protocol.
Collected from the article by \cite{tikader2022effects}.}	
\label{tau-R}
\end{center}
\end{figure}

Now, we have systematically deformed the geometrical shape of the square lattice and converted into rectangular structure by contracting its length along Y-axis $L_y$, while keeping $L_x$ fixed. In our experiment, Y axis-length of the lattice is varied from $L{_{y}} =$ 512 to $L{_{y}} =$ 2, keeping X-axis length fixed at $L{_{x}}=512$ and the instantaneous magnetisation ($m(t)$) is simultaneously measured for various system sizes. The dependence of the magnetic relaxation time on aspect ratio is critically investigated for the systems with PBC and OBC using two different choices of dynamical protocols. The ghaphical illustration of the outcomes are systematically shown in Figure-\ref{tau-R}. 

The relaxation time ($\tau$) is studied as a function of aspect ratio ($R$) in double-logrithmic space. The Figure-\ref{tau-R}(a) represents the obtained data, implementing Metropolis dynamics for the 2D systems under OBC, while Figure \ref{tau-R}(b) illustrates our observed results for PBC. Besides, our findings related to Glauber protocol are similarly demonstrated in the figure-\ref{tau-R}(c) and the figure-\ref{tau-R}(d) for OBC and PBC, in respective order. The resulting data collected for two temperatures above $T_c$, fixed at $T = 2.325~\rm J/k_B$ and $T = 2.4~\rm J/k_B$ are simultaneouly depicted in the Figure-\ref{tau-R}. We have encountered an interesting scenario. Across all cases, the relaxation time $\tau$ is found be almost independant of aspect ratio in the small $R$ region. On the contrary for higher values of aspect ratio, the system's relaxation time exhibits a strong dependency upon aspect ratio $R$. The graphical plots and fit statistics clearly imply that $\log(\tau)$ in the large $R$ limit, is a linear function of log($R$) with a negative slope ($-s$). The observation leads us to claim a power-law type relationship between relaxation time ($\tau$) and aspect ratio ($R$), i.e., $\tau \sim R^{-s}$. From further analysis we can firmly suggest that the exponent associated with the power-law relation ($s$) is a function of temperature as well. This temperature dependence property of the exponent `$s(T)$' does not change for any boundary condition and dynamics used. 

\begin{figure}[h!]
\begin{center}

\includegraphics[angle=0,scale=0.63]{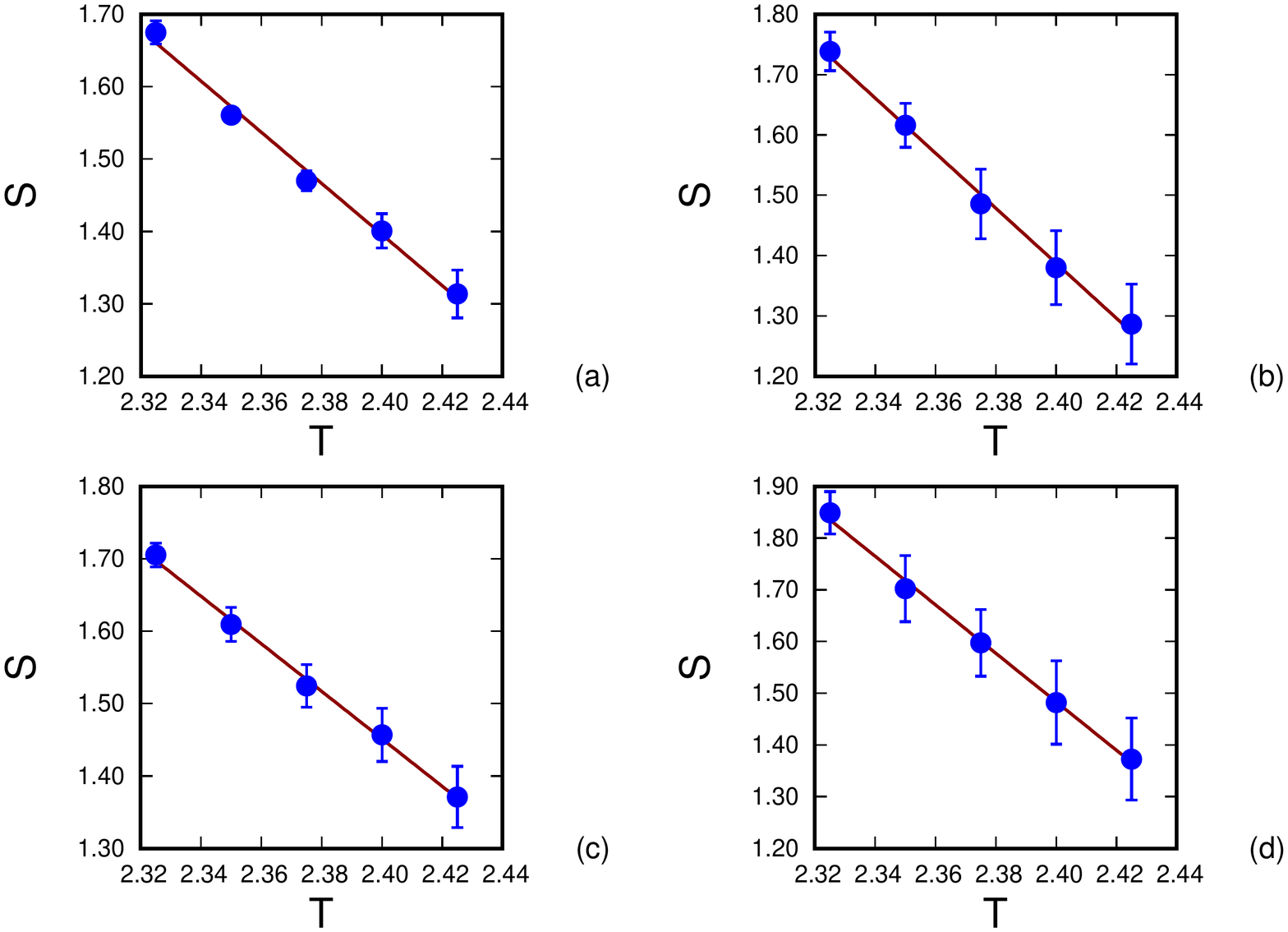}

\caption{Temperature variation of the exponent $s$ (`area non-conserving deformation') with linear fit; implementing
(a) OBC and Metropolis algorithm
(b) PBC and Metropolis algorithm
(c) OBC and Glauber algorithm and
(d) PBC and Glauber algorithm. This diagram is collected from the article  by \cite{tikader2022effects}.} 
\label{exponent-T}
\end{center}
\end{figure}

 In order to identify the precise temperature dependence of exponent in the power-law relation, a comprehensive analysis should be executed across a short range of temperatures near critical point ($T \rightarrow T_c$) in the paramagnetic regime. The resulting data are demonstrated in the Figure-\ref{exponent-T}, where the graphical figures \ref{exponent-T}(a) and \ref{exponent-T}(b) represent the graphical illustrations of data obtained by employing Metropolis dynamics for the systems under corresponding boundary conditions OBC and PBC. The outcomes, produced by using Glauber protocol are similarly represented in the Figure-\ref{exponent-T}(c) and the Figure-\ref{exponent-T}(d) related to OBC and PBC respectively. The linear decrease of exponent is firmly indentified as we increase the temperature. The dependence is expressed by the function $s(T)=aT+b$. The best-fitted parameters a and b and statistical $\chi^2$ are clearly presented in the Table-\ref{Table1}. Any type of dynamical rule namely, Metropolis and Glauber yield qualitatively similar outcomes. Moverover, it is duly noted that the rate of decrease in the exponent caused by temperature ($-{{ds} \over {dT}}$) of the system under PBC claims higher value compared to the system with OBC .
 
 The obtained results by implementing two types of boundary conditions and dynamical algorithms are provided in the Table-\ref{Table1}.
\begin{table}[h!]
\begin{center} {\bf{Table 1}} \\
 \vspace{0.2cm}   
\begin{tabular}{|p{2.5cm}|p{3.5cm}|p{2cm}|p{2cm}|p{2cm}|p{1cm}|}
  \hline
  Dynamical protocol & Boundary condition  & $a$ & $b$ & ${\chi }^2$  & DoF \\ 
  \hline 
   \multirow{2}{*}{Metropolis} & OBC & $-3.53\pm 0.18$ & $9.87\pm 0.42$ & 0.0006 &  3 \\ [1ex] \cline{2-6}
   & PBC & $-4.56\pm 0.17$ & $12.33\pm 0.41$ & 0.0005 & 3 \\ [1ex]
   \hline
   \multirow{2}{*}{Glauber} & OBC & $-3.28 \pm 0.11$ & $9.33 \pm 0.25$ & 0.0002 &  3 \\ [1ex] \cline{2-6} 
   & PBC & $-4.69 \pm 0.16$ & $12.74 \pm 0.38$ & 0.0004 & 3 \\ [1ex]
   \hline
  \end{tabular}
 \caption{ The best fitted values and statistical ${\chi }^2$  for linear fit of exponent $s$ versus temperature $T$ data set obtained in the context of `area non-conserving deformation'. Table is taken from the paper by \cite{tikader2022effects}.}
 \label{Table1}
\end{center}
\end{table}

\subsubsection{\bf{Rectangular geometry: Conserved area}}
\begin{figure}[h!]
\begin{center}
\includegraphics[angle=0,scale=0.63]{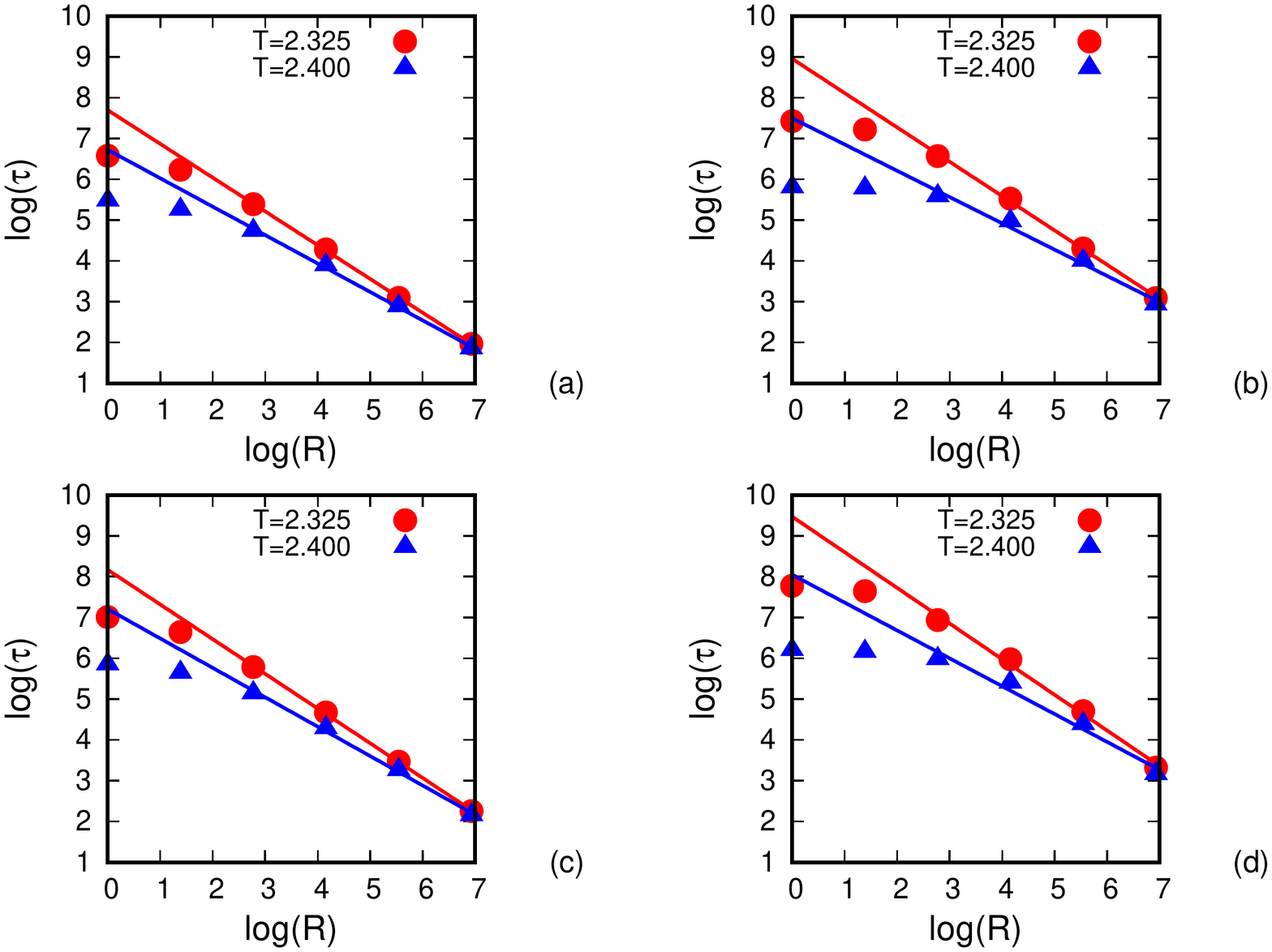}
\caption{Geometrical dependence of the relaxation time. Variation of $\tau$ 
with aspect ratio in a double-logarithmic space. Red circles symbolize the data obtained at $T=2.325~J/k_B$ and Blue triangles indicate the temperature $T=2.4~J/k_B$ in the context of `area-conserving deformation'. Linear fit ($\log(\tau)=-s\times \log(R)+r$) is indicated by Red line for $T=2.325~J/k_B$ and Blue line for $T=2.4~J/k_B$. Results obtained using
(a) OBC and Metropolis dynamics, 
(b) PBC and Metropolis dynamics,
(c) OBC and Glauber protocol and
(d) PBC and Glauber protocol.
The diagram is taken from the article by \cite{tikader2023effects}.}
\label{tau-R1}
\end{center}
\end{figure}

What consequences would arise, if we reshape the square lattice into rectangular form while ensuring that the overall area of the lattice remains unchanged and gradually defomed in a quasi-one dimensional strip? How does this area-conserving deformation affect the relaxation time in ferromagnetic monolayer? This is an intriguing question that demands our attention. The geometrical shape of the system is systematically changed by transforming the square shaped lattice into a rectangle, while conserving the area of the lattice $L_x \times L_y$. The temporal variation of magnetisation $m(t)$ is minutely studied for the area-conserved rectangular geometry. The length along X-axis is chosen from $L_x=64$ to $L_x = 2048$ and the corresponding length along Y-axis  $L_y$ is changed as well, ensuring the area-conservation at fixed value $L_x \times L_y$=4096. The variation of relaxation time caused by geometrialc deformations is critically studied for the systems with OBC and PBC employing the different algorithms (Metropolis dynamical rule and Glauber protocol). The figure-\ref{tau-R1} illustrates how log values of relaxation time, $\log\tau$ explicitly depends on the logarithm of aspect ratio ($\log R$). The systematic analysis reveals that in the limit of large aspect ratio ($R \gg 1$) the dependence of relaxation time upon aspect ratio can be expressed by the power-law type function, i.e., $\tau \sim R^{-s}$. Interestingly, this behavior closely resembles the case of `area non-conserved deformation'. A notable thermal dependence of the exponent $s$ associated with power-law relation is reported as well.

\begin{figure}[h!]
\begin{center}
\includegraphics[angle=0,scale=0.63]{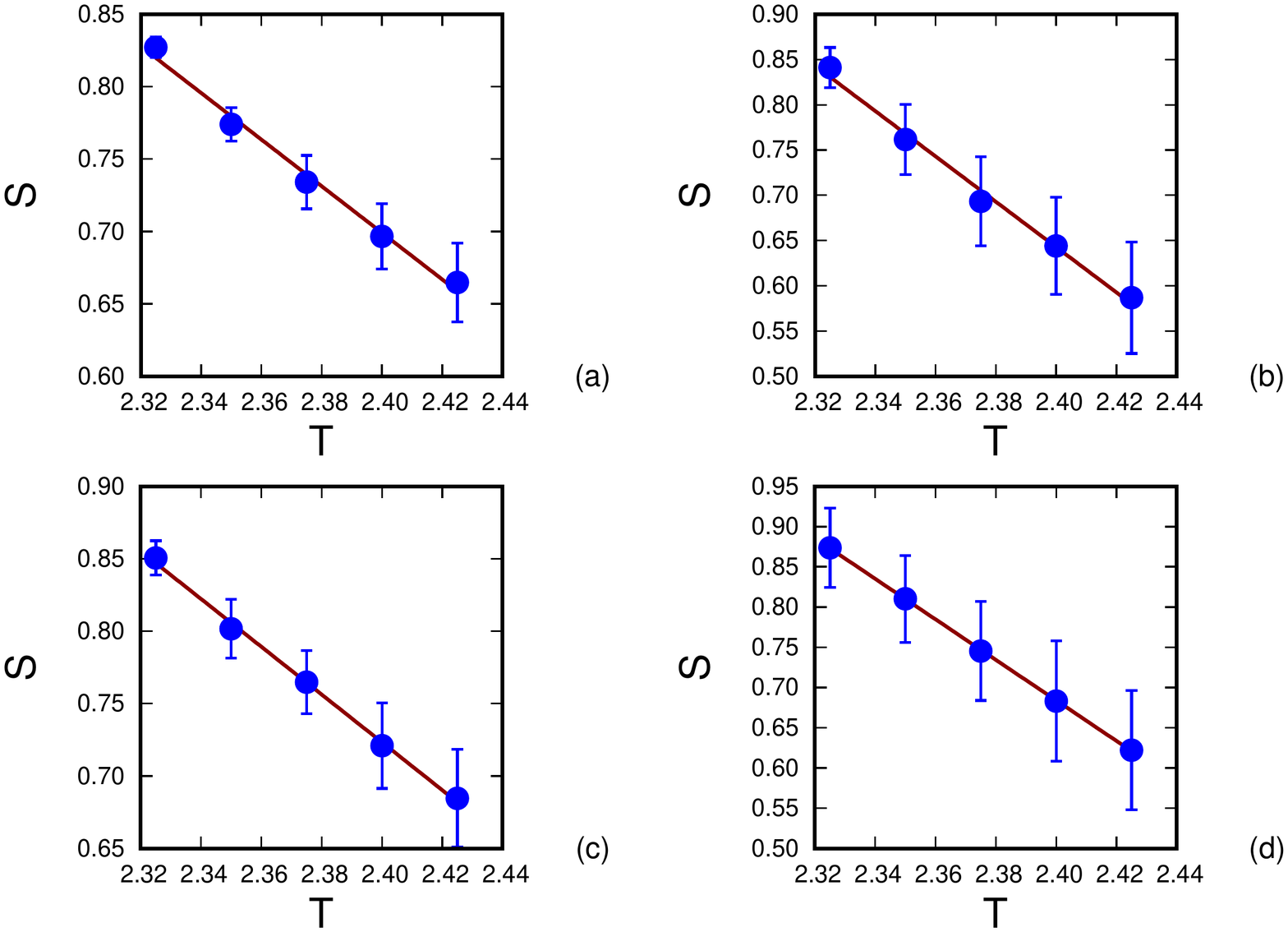}
\caption{Temperature variation of the exponent $s$ (`area-conserving deformation') with linear fit; implementing
(a) OBC and Metropolis dynamics
(b) PBC and Metropolis dynamics
(c) OBC and Glauber protocol and
(d) PBC and Glauber protocol.
\cite{tikader2023effects}.}
\label{exponent-T1}
\end{center}
\end{figure}

Our motivations lead us to explore the thermal dependence of the exponent $s$ and elucidate such temperature-dependant behavior. The values of the exponent $s$ related to power-law function are estimated for a short-range of temperature near critical point in the region above $T_c$ and the thermal variation is represented in figure-\ref{exponent-T1}. Strikingly similar behavior is exhibited here. The linear decrease in exponent with temperature is firmly noted. So, the temperature-dependant behaviour of $s$ can be effectively discribed by a linear function with negetive slope. The straight-line function is  $s(T)=aT+b$ ($a<0$). The best-fitted parameters a and b along with statistical ${\chi}^2$ are demonstrated in the Table-\ref{Table2} for different dynamical rules and boundary conditions.
\begin{table}[h!]
\begin{center} {\bf{Table 2}} \\
 \vspace{0.2cm}   
\begin{tabular}{|p{2cm}|p{3.5cm}|p{2cm}|p{2cm}|p{2cm}|p{1cm}|}
  \hline
  Dynamical rule & Boundary condition  & $a$ & $b$ & ${\chi }^2$  & DoF  \\ [1ex]
  \hline 
   \multirow{2}{*}{Metropolis} & OBC & $-1.61\pm 0.09$ & $4.56\pm 0.22$ & 0.0001 &  3 \\ [1ex] \cline{2-6}
   & PBC & $-2.51\pm 0.14$ & $6.66\pm 0.32$ & 0.0003 & 3 \\ [1ex]
   \hline
   \multirow{2}{*}{Glauber} & OBC & $-1.65\pm 0.05$ & $4.68\pm 0.11$ &  $4.1\times 10^{-5} $&  3 \\ [1ex] \cline{2-6} 
   & PBC & $-2.52\pm 0.02$ & $6.73\pm 0.04 $ & $5.2\times 10^{-6}$ & 3 \\ [1ex]
   \hline
  \end{tabular}
 \caption{ The best fitted values and statistical ${\chi }^2$  for linear fit of exponent $s$ versus temperature $T$ data set obtained in the context of `area-conserving deformation'. \cite{tikader2023effects}}
 \label{Table2}
\end{center}
\end{table}
\subsubsection{\bf {Dynamical evolution of Spin-flip density}}
The spin-flip density holds a great significance to study the relaxation dynamics in the Ising system, as it effectively indicates how fast a thermodynamic system attains the state of thermal equillibrium. By the defination, spin-flip density is formulated as follows,
\begin{equation}
    f_d(t)=\frac{\text{\rm The~no.~of~spin-flips~at~$t^{\rm th}$~MC Step per Spin}}{\text{\rm Total~no.~of~lattice~sites}}
 \end{equation} 
 A 2D system with square lattice of size $L$ is considered. The total spin or site counted in the lattice $N = L^2$, while $N_{\rm surface} = (4L-4)$ no. of spins at surface and $N_{\rm core} = (L-2)^2$ number of sites in the core. The evolution of spin-flip $f_d(t)$ density with time is thoroughly studied. The observation is duly note for surface and core spin simultaneously. Starting from the all spins up configuration, the two dimensional system gradually relax towards the equilibrium state. As a result the spin-flip density continues to evolve with time and in the course of time it saturates at some fixed values after attaining the original equilibrium. The saturated value of spin-flip  density $f_{sd}$ for surface as well as core are reported to vary with the system size $L$.\\
\begin{figure}[h!]
\begin{center}
\includegraphics[angle=0,width=0.45\textwidth]{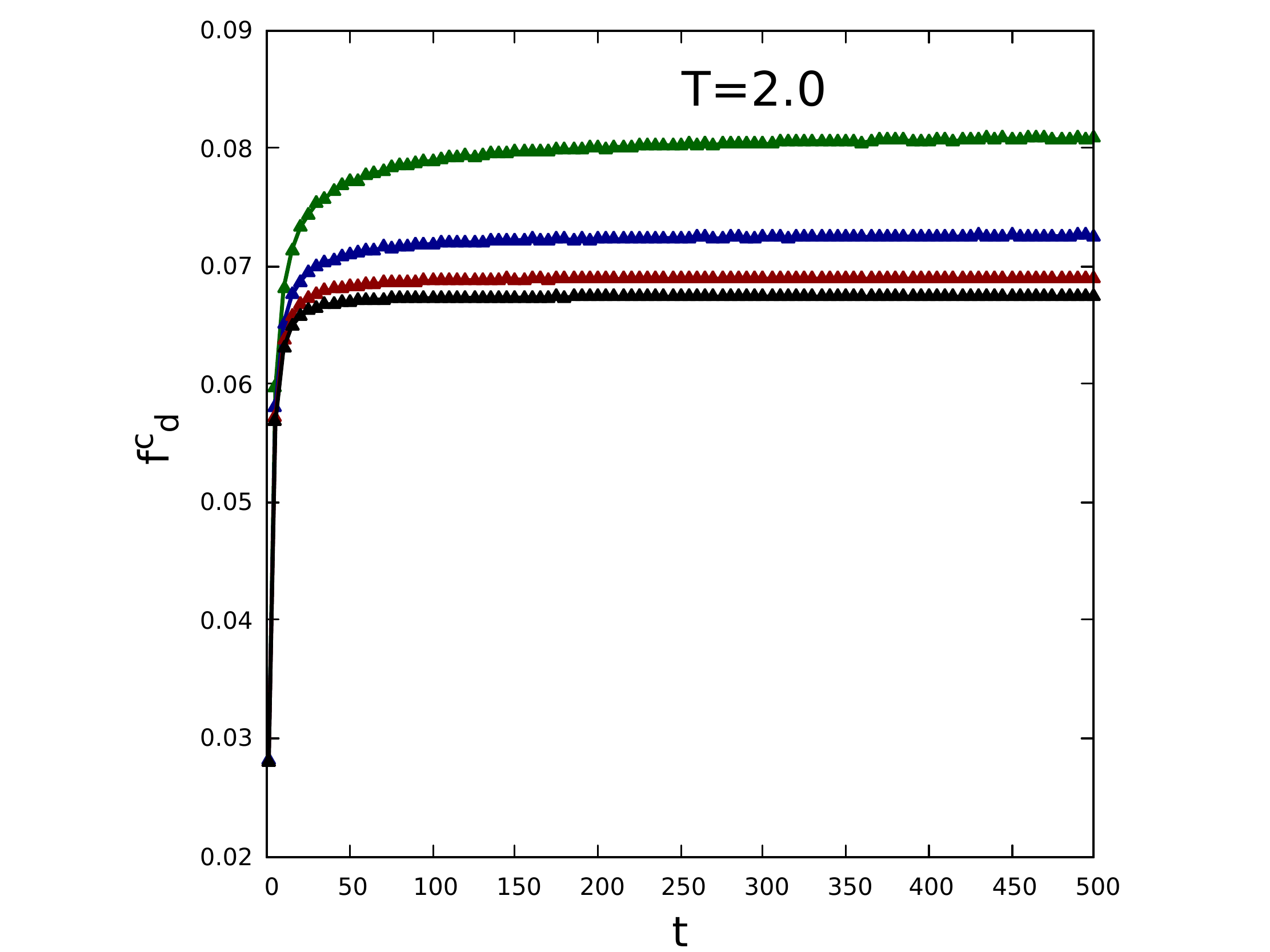}(a)
\includegraphics[angle=0,width=0.45\textwidth]{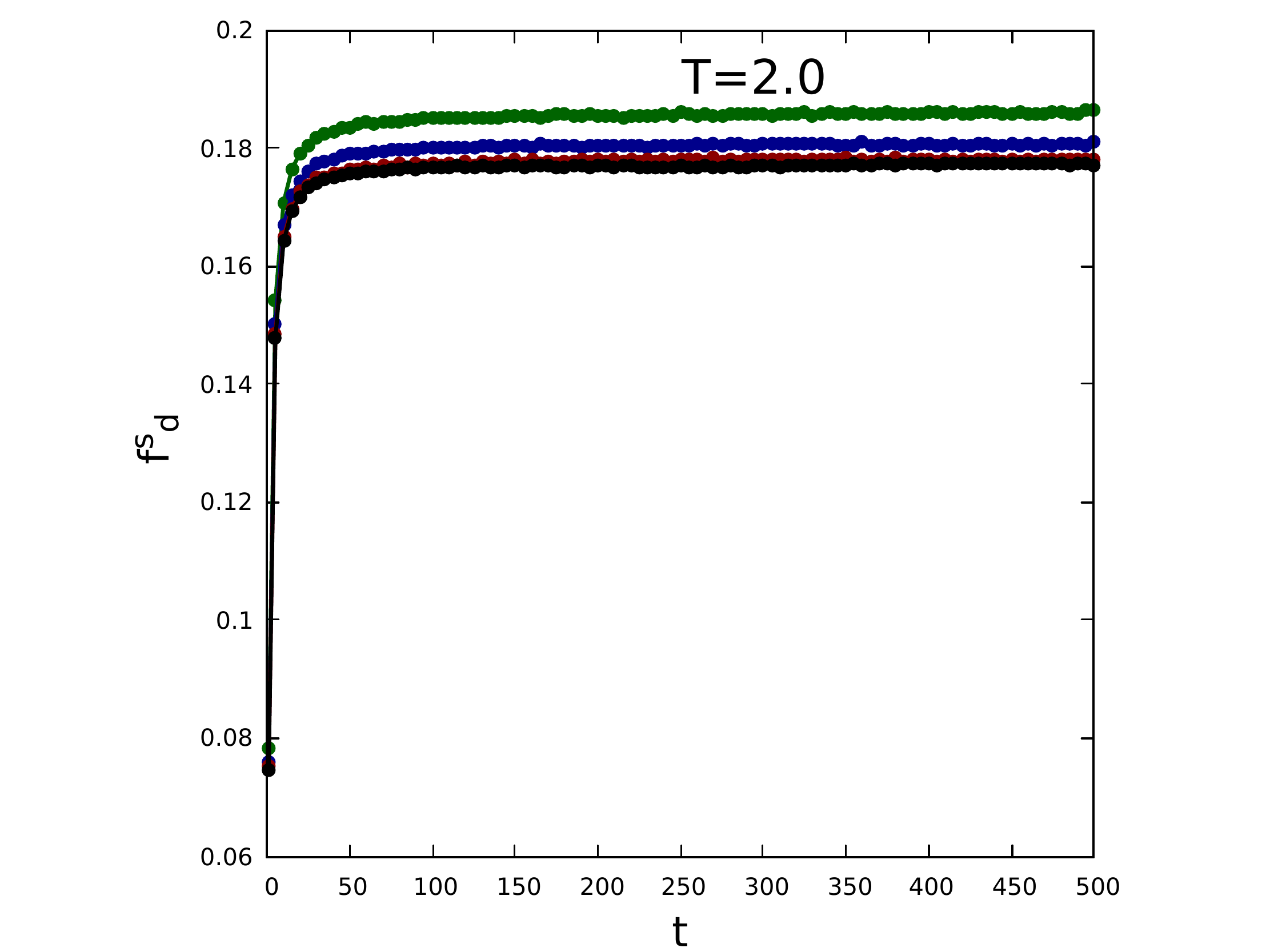}(b)
\\
\includegraphics[angle=0,width=0.45\textwidth]{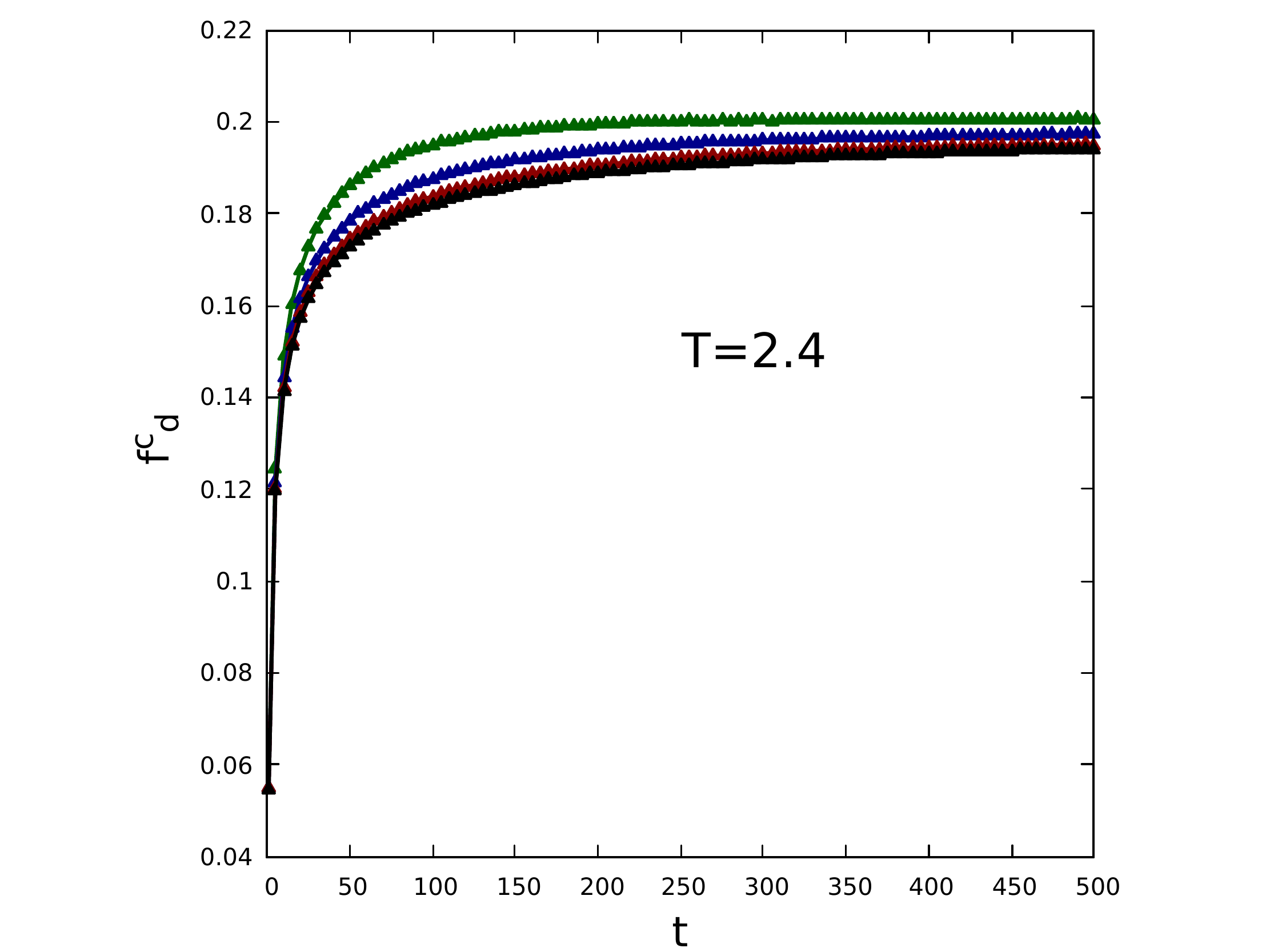}(c)
\includegraphics[angle=0,width=0.45\textwidth]{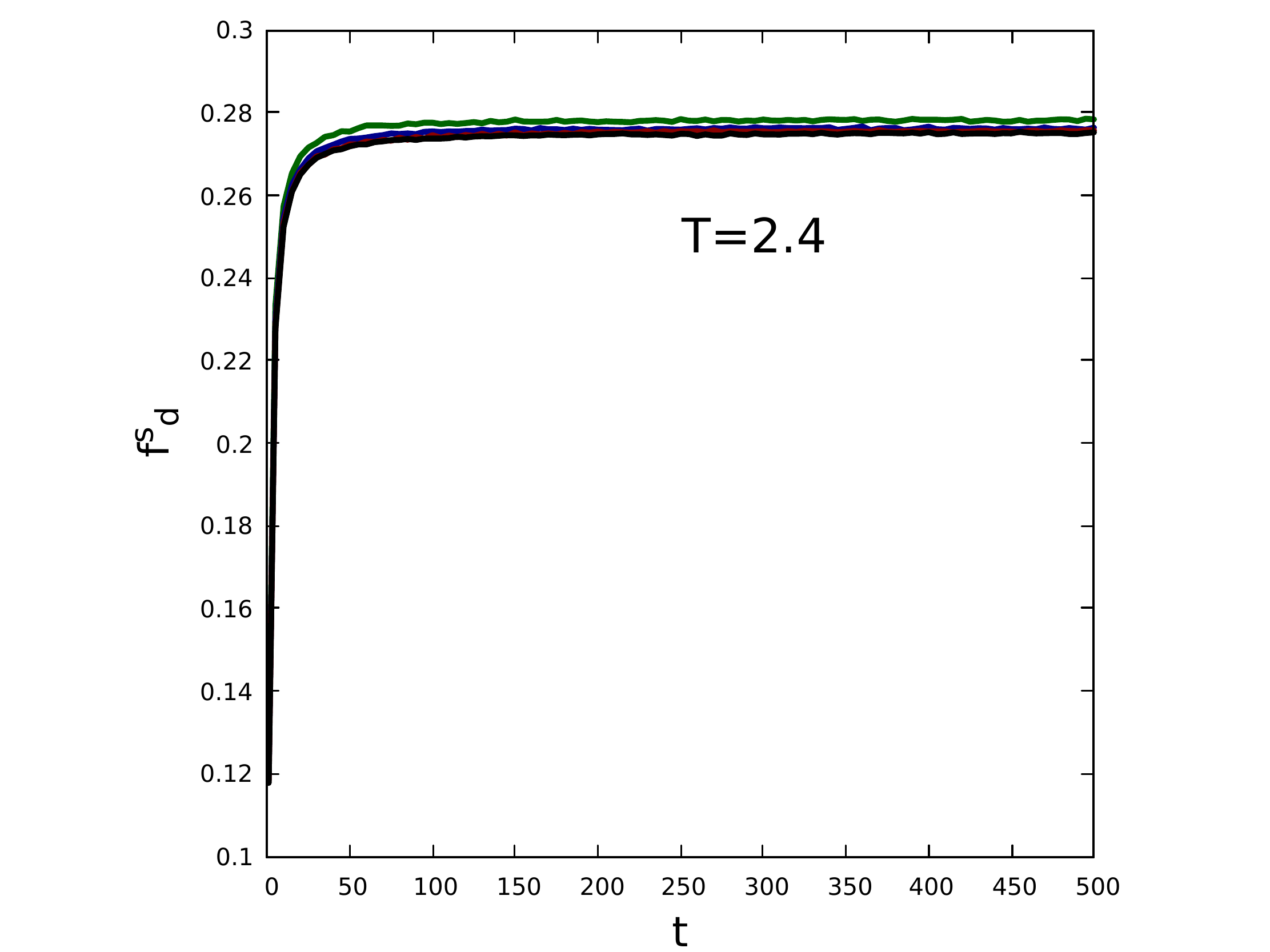}(d)
\caption{Time evolution of the spin-flip density of the core of the lattice ($ f^c_d $) at (a)$T=2.00~J/k_B$ (Ferromagnetic region) and (c) $T=2.40~J/k_B$ (Paramagnetic region) for lattice size; $L = 25$ (green), 50 (blue), 100 (red) and 200 (black). Time evolution of spin-flip density for surface ($f^s_d$) with time at (b) $T=2.00~J/k_B$ and (d) $T=2.40~J/k_B$. Lattice size: $L = 25$ (green), 50 (blue), 100 (red) and 200 (black). Data produced by the Glauber protocol. Taken from the article by \cite{tikader2023effects}.}
\label{fdg}
\end{center}
\end{figure}
  It should noted that each spin of the core part of the lattice can interact with four nearest-neighbouring spins, whereas spin located on the edge interacts with three nearest-neighbouring spins and each spin at four corner of the lattice can experience the interactions with only two nearest-neighbours. To investigate surface relaxation, it is necessary to examine the system with unbounded surfaces, thereby rendering the open boundary condition as only applicable choice.

\begin{figure}[h!]
\begin{center}

\includegraphics[angle=0,width=0.45\textwidth]{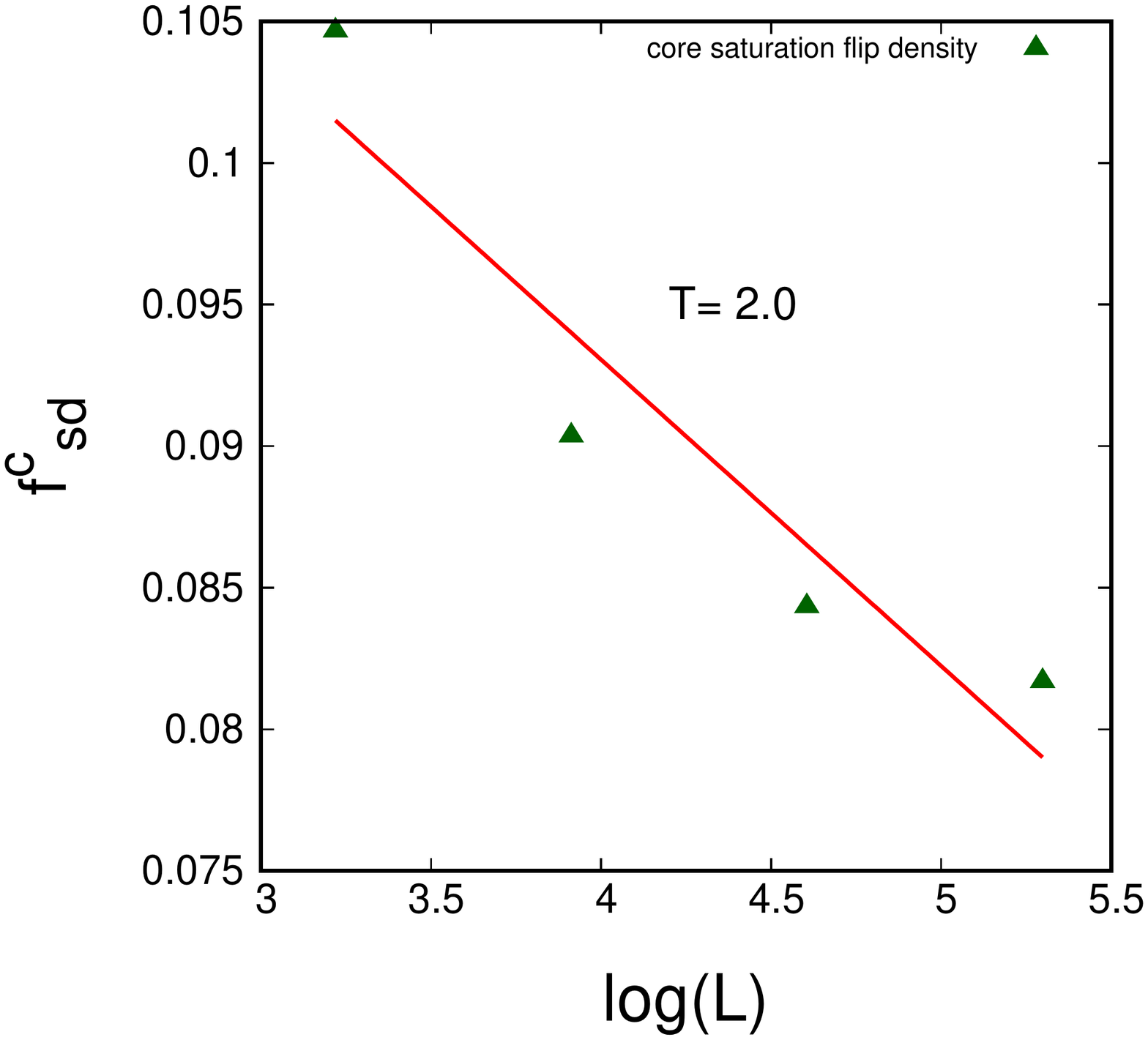} (a)
\includegraphics[angle=0,width=0.45\textwidth]{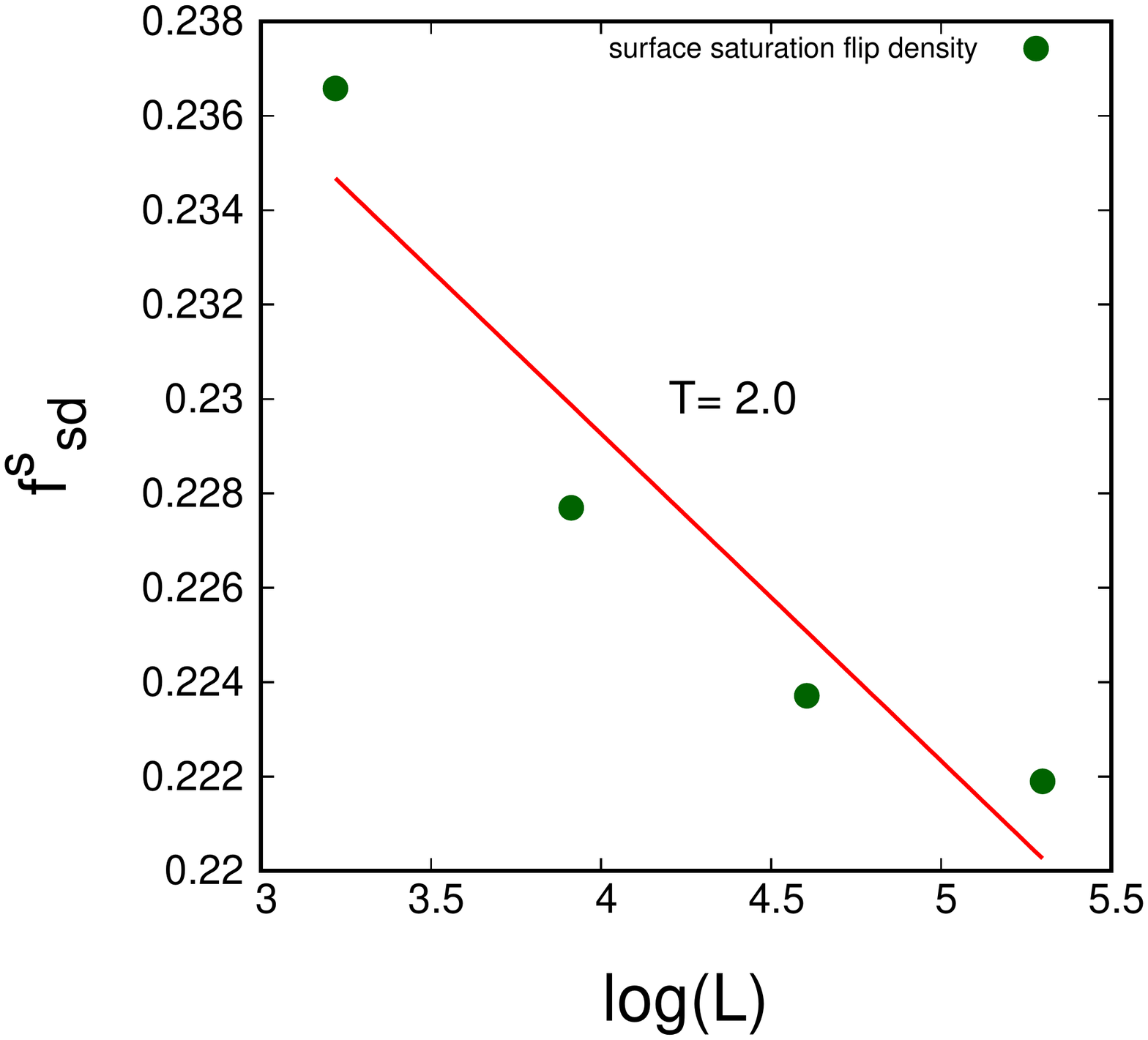} (b)
\\
\includegraphics[angle=0,width=0.45\textwidth]{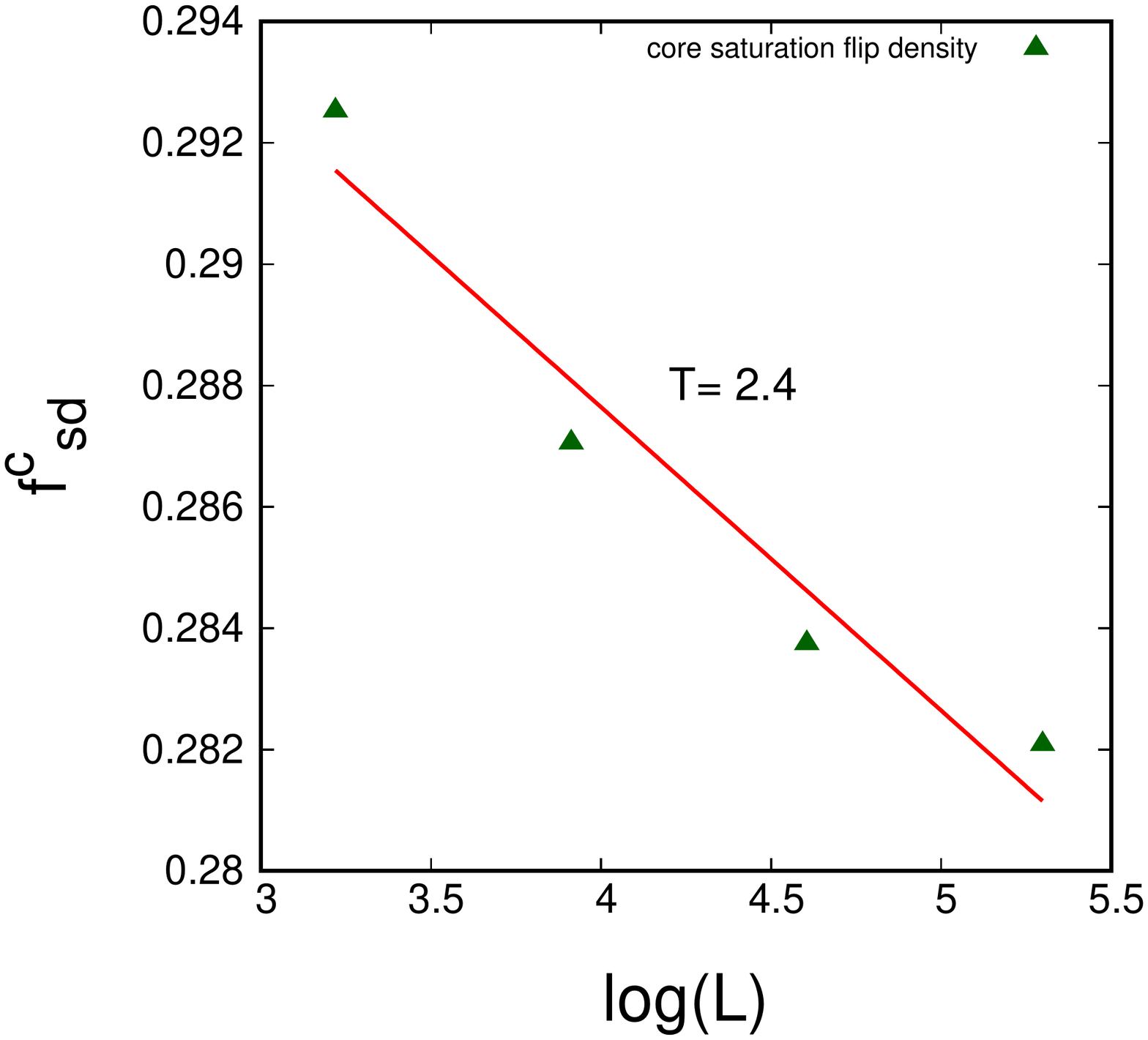} (c)
\includegraphics[angle=0,width=0.45\textwidth]{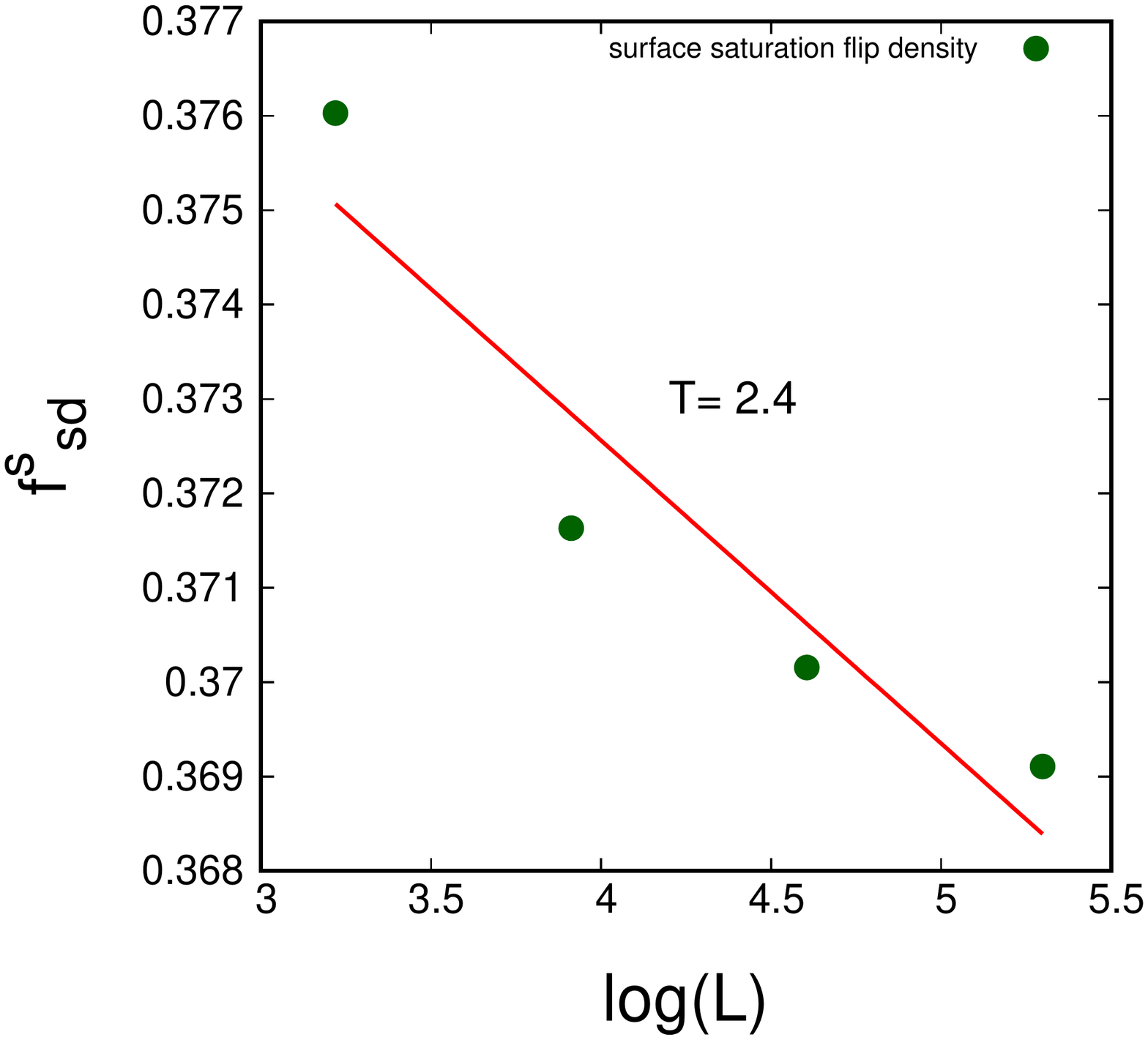} (d)
\caption{The variation of spin-flip density after saturation  with logarithm of size. The best linear fit is graphically illustrated for core ($f^c_{sd}$) and surface ($f^s_{sd}$) at $T=2.00~J/k_B$ (Ordered phase) and $T=2.40~J/k_B$ (Disordered phase). Resulting data are obtained under the Metropolis protocol. The triangle and bullet symbols indicate core and suface respectively. This diagram is taken from the article by \cite{tikader2023effects}.}
\label{fdlfitmetro}
\end{center}
\end{figure}
 The spin-flip density is determined by measuring the sample-average over a large number of different uncorrelated samples. The simulation is performed in both temperature region below $T_c$ or ferro-phase ($T=2.0 ~J/k_B$) and above $T_c$ or para-phase ($T=2.4 ~J/k_B$). The figure-\ref{fdg}(a) depicts the temporal dependence of the spin-flip density in the core $f^c_d$, where the figure-\ref{fdg}(b) represents that in the surface $f_d^s$, fixed at $T=2.0~J/k_B$ implementing the Glauber protocols. It is prominent that the  spin-flip density of the core as well as surface eventually approaches the saturation region after a spicific time-duration by achieving relaxation. The simulation is conducted for various lattice sizes from $L$ = 25 to 200 revealing intriguing variations in the saturation values based on the system size $L$. Interestingly, the saturated value of spin-flip density in the surface ($f_d^s$) surpasses the saturated value for the core ($f^c_d$). The scenario is noted for any given value of system size $L$. The figure-\ref{fdg}(c) and figure-\ref{fdg}(d) illustrate the changes in the spin-flip density over time, in the context of core and surface part in due order, at the temperature $T = 2.4~J/k_B$. The spin-flip density after saturation $f_{sd}$ is noted be relatively higher valued in the disordered phase (para-phase $T=2.4~J/k_B$) for any given size. Through the investigations, an interesting trend is explored for surface and core part of lattice, where the saturated value of spin-flip density decreases as the system size ($L$) increases, both in the ferromagnetic regime and the paramagnetic regime.

The aforementioned consecutive simulation steps are repeated applying the Metropolis dynamics, yielding qualitatively similartype of results to those obtained previously with Glauber dynamics. Neverthless, the result remarks that the Metropolis algorithm assures strikingly higher values of saturated spin-flip density compared to Glauber protocol.
\\
The saturated value of spin-flip density ($f^c_{sd}$, $f^s_{sd}$) is determined by taking the time average of the spin-flip density over the last 100 MC Step per Spin. The dependence of saturated spin-flip density $f_{sd}$ on the size ($L$) of the system is critically analyzed. The observations indicate that the saturated spin-flip density ($f^c_{sd}$, $f^s_{sd}$) linearly varies with logarithm of size $\log(L)$, i.e., $f_{sd}=a + b~\log(L)$ for both core and surface spin. Strikingly similar outcomes are found for both Metropolis and Glauber protocols. The exploration is carried out in both ferromagnetic phase ($T=2.0$) and paramagnetic phase ($T=2.4$). 
Here, complete results are depicted in the graphical figure-\ref{fdlfitmetro}.
 Table-\ref{Table3} and Table-\ref{Table4} provide the best-fitted parameters and statistical ${\chi}^2$ to indicate goodness of fit for the results found by employing Glauber and Metropolis algoritms respectively.
  
\begin{table}[h!]
\begin{center} {\bf{Table 3}} \\
 \vspace{0.2cm}   
\begin{tabular}{|p{3.0cm}|p{2.5cm}|p{2.5cm}|p{3cm}|p{2cm}|p{1cm}|}
  \hline
  Temperature region  & Part of Lattice & a & b & ${\chi }^2$  & DoF \\ [1ex]
  \hline 
   \multirow{2}{*}{Ferro phase (2.0)} & Core & $0.099\pm0.006$& $-0.006\pm0.001$ & 0.000012 &  2 \\ [1ex] \cline{2-6}
   & Surface & $0.198\pm 0.005$ & $-0.004\pm 0.001$ & 0.000005 & 2 \\ [1ex]
   \hline
   \multirow{2}{*}{Para phase (2.4)} & Core & $0.210 \pm 0.002$ & $-0.003 \pm 0.0005 $ & 0.000001 &  2 \\ [1ex] \cline{2-6} 
   & Surface & $0.282\pm 0.001$ & $-0.001\pm 0.0003$ & 0.000007 & 2 \\ [1ex]
   \hline
  \end{tabular}
 \caption{The best fitted values of linear fit for  $f_{sd}$ vs. $\log L$ data;  employing Glauber Dynamics. Collected from \cite{tikader2023effects}. }
 \label{Table3}
\end{center}
\end{table}
\begin{table}[h!]
\begin{center} {\bf{Table 4}} \\
 \vspace{0.2cm}   
\begin{tabular}{|p{3cm}|p{2.5cm}|p{2.5cm}|p{3cm}|p{2cm}|p{1cm}|}
  \hline
  Temperature region & Part of Lattice & a & b & ${\chi }^2$  & DoF \\ [1ex]
  \hline 
   \multirow{2}{*}{Ferro phase 2.0} & Core & $0.136\pm0.012$& $-0.011\pm0.003 $ & 0.000035 &  2 \\ [1ex] \cline{2-6}
   & Surface & $0.257\pm 0.007$ & $-0.007\pm 0.001$ & 0.000013 & 2 \\ [1ex]
   \hline
   \multirow{2}{*}{Para phase 2.4} & Core & $0.308\pm 0.004$ & $-0.0050 \pm 0.0009 $ & 0.000003 &  2 \\ [1ex] \cline{2-6} 
   & Surface & $0.385\pm 0.003$ & $-0.0032\pm 0.0008$ & 0.000003 & 2 \\ [1ex]
   \hline
  \end{tabular}
 \caption{The best fitted values of linear fit for  $f_{sd}$ vs. $\log L$ data; employing Metropolis Dynamics. Collected from \cite{tikader2023effects}}
 \label{Table4}
\end{center}
\end{table}



%
\vskip 1cm
\section {Magnetic relaxation: Experimental data}

\subsection{\bf Slow relaxation near transition point }
\emph{Critical slowing down (CSD)} in the relaxation process of order parameter (or divergence of time scale at transition point) is crucial in understanding the  dynamicals of second-order phase transition and critical phenomena (Eq.\ref{eqn:18}). The slowing down of relaxation time arises due to the presence of long-range correlations and large fluctuations in order parameter near critical point. 

The correlation length is a measure of the spatial extent over which fluctuations in the system are correlated. Slowing down of the relaxation time is strongly related to the divergence of the correlation length at critical point. The relaxation time $\tau$  is a function of correlation length ($\xi$) and $\xi$ also exhibits a power-law dicritical point. So we can establish the power-law relation between relaxation time and temperature.  (Goldenfeld 1992)\cite{Goldenfeld:92}

\begin{equation}
    \tau (T) = {\tau_0} \left[ \dfrac {\xi (T)}{\xi_0} \right]^z = \tau_0 (\varepsilon^{-\nu})^z
\label{eqn:24}
\end{equation}
Here $z$ is the dynamic critical exponent and $\nu$ is defined as the critical exponent assiciated with correlation length ($\nu=1.0$ in two dimensional Ising Universality class). $\varepsilon$ reprents the reduced temperature, which is termed as $\varepsilon = {{T-T_c}\over T_c}$ above $T_c$.  

\begin{figure}[h!]
\begin{center}
\includegraphics[angle=0, scale = 0.4]{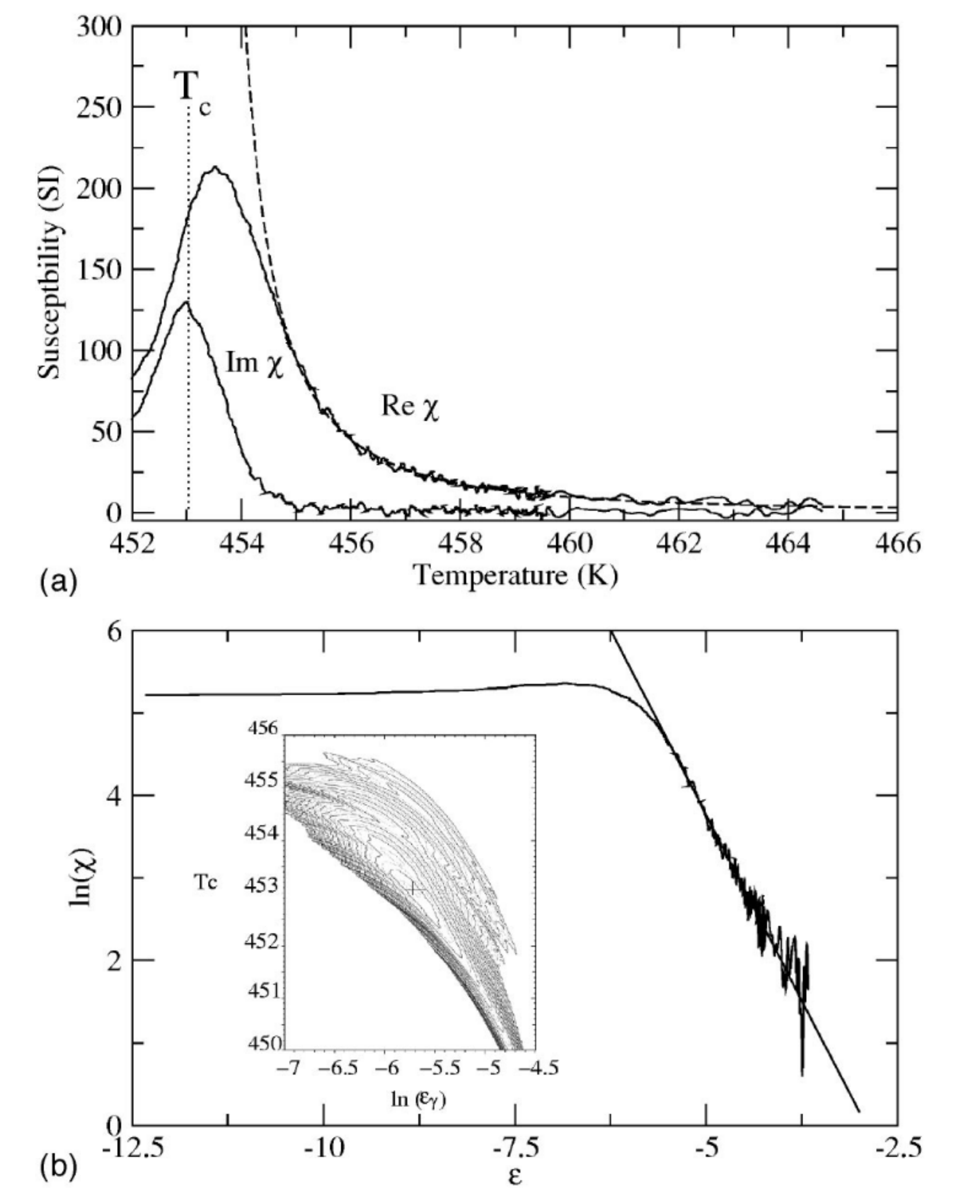} 
\caption{(a)The both real and imaginary parts of the complex susceptibility as the function of temperature (in unit of K) measured for a two monolayer of Fe on W(110). The dashed line provides the power-law type fitting for the real component of susceptibility to estimate the critical temperature and static critical exponent of susceptibility $\gamma$. Here dotted line indicates the position of critical temperature i.e., $T_c = 453.03$K. (b) The logarithm of real component of susceptibility is plotted against the log of reduced temperature $\ln(\varepsilon)$ with the line fit shown by the solid line.   The variance of the fit (i.e., reduced statistical $\chi^2$) as a function of critical temperature $T_c$ and cut off $\epsilon_{\gamma}$ is graphically illustrated in a contour plot (shown in inset figure). The minimum point of varience identified by the cross symbol, gives the best values of $ T_c $ and $ \varepsilon_\gamma $. These graphical plots are collected from the paper authored by \cite{PhysRevB.71.144406}}
\label{sucep}
\end{center}
\end{figure}

Slow relaxation process (CSD) in the ferromagnetic monolayer near critical point is experimentally studied by examining the ultrathin film of Fe/W(110) and critical properties are meticulously investigated in the research article by Dunlavy and Venus, 2005. The real and imaginary components of the complex valued AC magnetic susceptibility (ACMS) are critically measured for the two monolayers of Iron(Fe) deposited on W(110) surface. The SMOKE (surface magneto-optic Kerr effect) is used here along with Lock-in technique to study complex suceptibility and other magnetic properties \cite{PhysRevB.71.144406}.

The significant statistical analysis provides the best values of critical temperature for Fe/W(110) film, i.e.,  $\rm T_c = 453.03\pm 0.02$ K. The observation is mentioned in the graphical figure-\ref{sucep}(a). 

\begin{figure}[h!]
\begin{center}
\includegraphics[angle=0, scale = 0.35] {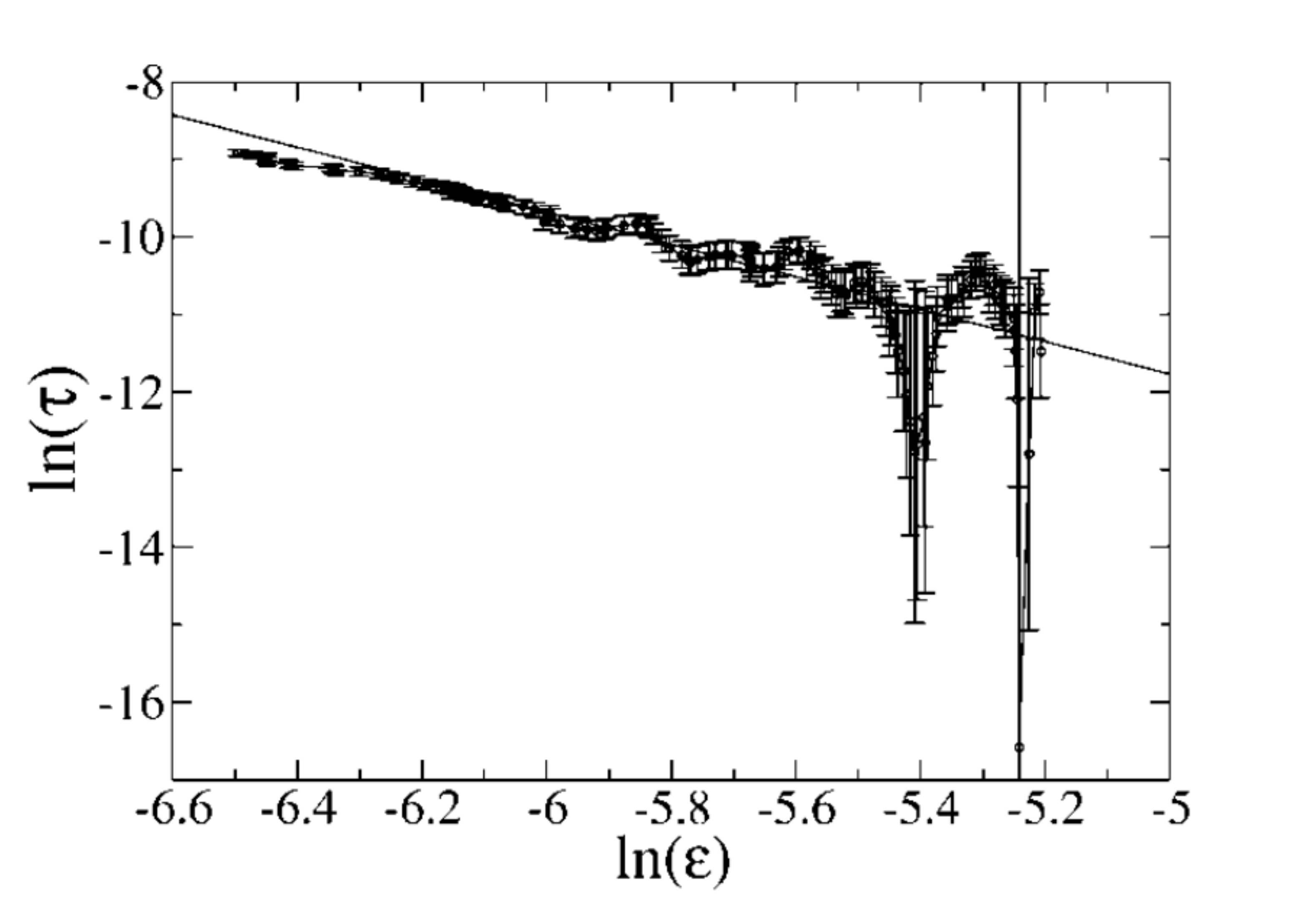} 
(a)
\\
\includegraphics[angle=0, scale = 0.35] {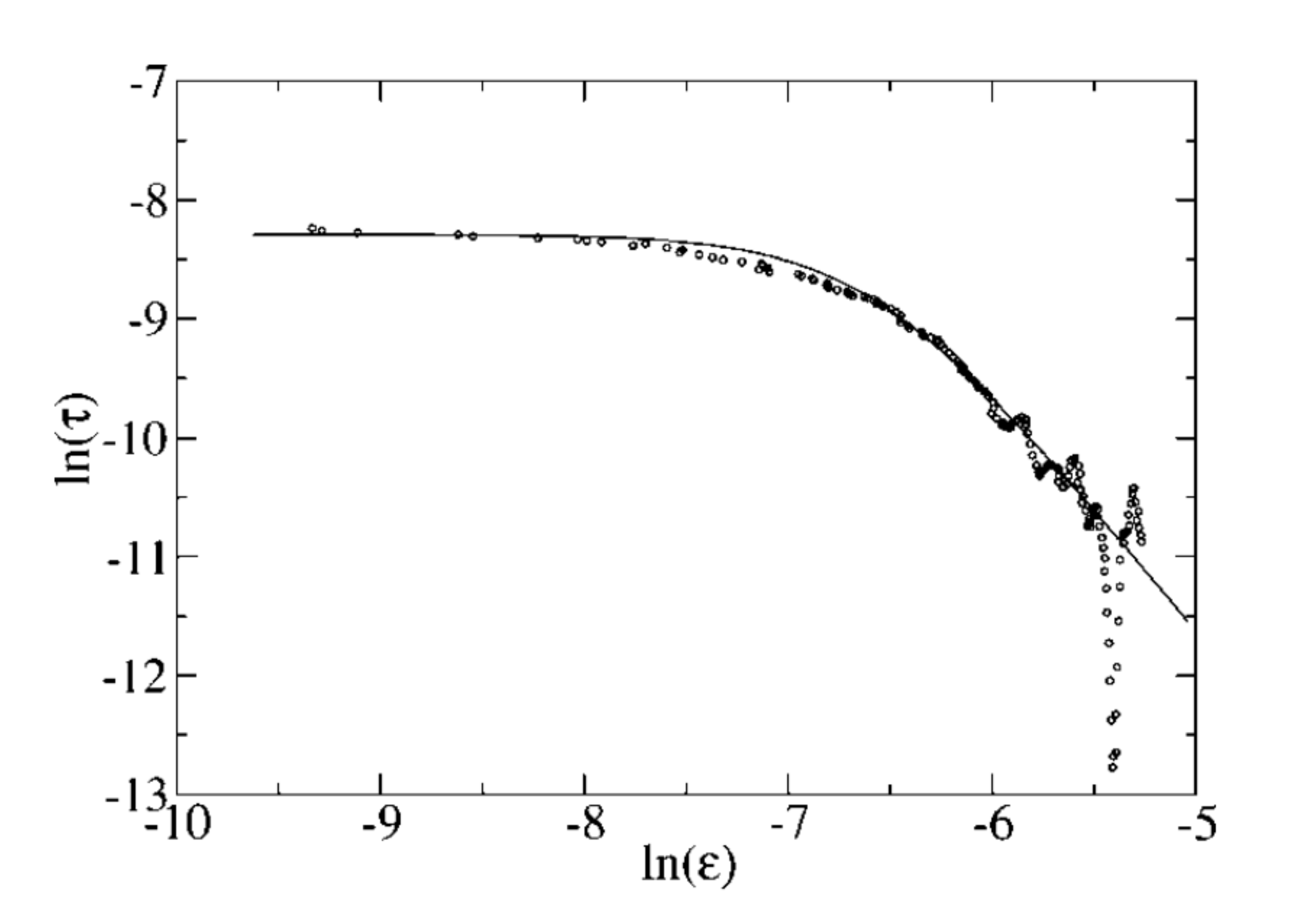} 
(b)

\caption{(a) The variation of relaxation time ($\tau$) with reduced temperature ($\varepsilon = {{T-T_c}\over T_c}$) in double-logarithmic space. Here the linear fit is demonstrated for a finite range of temperature considering $\ln{(\varepsilon)}$ values from $-6.51$ to $-5.26$. (b) Dependences of relaxation time ($\tau$) on reduced temperature $\varepsilon$ in double-log space. The solid line represents the best fitting for saturated correlation length $\xi_{sat}$ using a model that incorporates both saturation regions and power-law type scaling. Collected from \cite{PhysRevB.71.144406}.}
\label{relaxation_time}
\end{center}
\end{figure}

The magnetic relaxation time is estimated by taking the ratio of imaginary and real component of the complex susceptibility, i.e., $\tau (T) = 
\dfrac {\rm Im [\chi (\textit{T})]} {\omega ~ \rm Re[\chi (\textit{T})]} $  and its variation with reduced temperature $\varepsilon$ is graphically demonstrated in the double-logarithmic space (Figure \ref{relaxation_time}(a)). The power-law type behavior of $\tau$ is evident from the linear fit shown in the figure-\ref{relaxation_time}(a). By critical analyzing they have determined the best value of the dynamical critical exponent $z$ i.e. $ z=2.09\pm0.06$ within $95 \% $ confidence interval and also estimated critical exponent related to magnetic suscectibility $\gamma = 1.76 \pm 0.06$. The estimated value of $z$ confirms the consistency with the theoretical \cite{PhysRevLett.47.1837} and computational \cite{PhysRevLett.76.4548} studies that value of dynamic critical exponent $z$ (lies between 2.1 and 2.25) has lower bound 2 ($z\geqslant 2$) for the 2D Ising Universality class.

However, this power-law scaling of $\tau$ is disrupted after certain range because of the experimental limilations. The power-law scaling only extends closer to critical point $T_c$  over a certain range and beyond that limit $\tau$ eventually saturates at $ \varepsilon_z $ due to finite-size effect as well as finite-field effect. The result is depicted in the figure-\ref{relaxation_time}(b). 


\vskip 1cm

\section {Magnetic monolayer: Application and preparation:}

Magnetism and its wide applications in spintronics have revolutionised the development of high-density data storage solutions. The hard storage drive with high efficiency, is mainly prepared by exploiting the phenomena like giant magnetoresistance (GMR) and tunnel magnetoresistance (TMR). Moreover, the advent of fundamental mechanisms namely spin-transfer torque (STT) and spin-orbit torque (SOT) has propelled the advancment of spintronics technologies towrads the creation of highly efficient, non-volatile and ultra-low power memory circuits with high efficiency, e.g., magnetic random access memories (MRAMs). Now these groundbraking technologies have successfully integrated with CMOS-based devices fostering the growth of environment-friendly sustainable technology. The future of  spintronics holds immense potential. It offers exciting possibilities to combine logic and memory functions with computing architectures using spin variable. Additionally, numerous disruptive approaches including logic-in-memory computing, stochastic computing,  neuro-morphic computing, and quantum circuits are actively being explored to develop the post-CMOS era of computing \cite{D0NR07867K}.

Let us briefly mention here the preparatory methodology of magnetic monolayer (2D materials). The previous studies investigating the properties of two dimensional materials for spintronics are predominantly based on atomically thin layers, obtained through mechanical exfoliation from the bulk crystal. This technique is considered as an effective method for isolating the high-quality, atomically thin monolayer (two dimensional materials). However, this exfoliation technique has inherent limitations in terms of scalability. Since the exfoliation process conducted in an oxidative environment poses additional drawbacks, it is not suitable for spintronics related applications. As a result, there are recurring limitations when it comes to the integration and study of magnetic monolayer (two dimensional material) on conventional ferromagnetic spin sourcing materials, e.g.,Fe, Ni, Co and their alloys. In reality, the surface oxidation significantly hinders the spintronics properties of these materials, resulting in restricted spin extraction and compromised performances of transport in spin-valve effect. 

To overcome the obstacles associated with scalability and the integration of sensitive spintronics materials, the direct growth of two-dimensional materials along with exfoliation method can be highly beneficial. It is significantly observed (Och et. al, 2021) that for two dimensional (2D) material, e.g., h-BN and graphene, the usage of direct chemical vapour deposition (CVD) growth, becomes a key factor in the exposing of their spintronics properties.

\vskip 1cm

\section {Concluding notes:}

In the present article, we have represented a comprehensive analysis of our recently published study \cite{tikader2023effects} which focuses on Monte Carlo simulation-based investigation of the magnetic relaxation process in the 2D Ising ferromagnet. The study primarily addressed the impact of system's geometrical shape, applied boundary conditions and dynamical algorithms on the relaxation behaviour exhibited by ferromagnetic Ising model. The major findings are highlighted here. 

The temporal decay of the magnetisation evidently follow the exponential law, $m(t) \thicksim \exp (-{t \over {\tau}})$. It is remarkable that the magnetic relaxation time has strong dependency  upon the aspect ratio ($R = {{L_x}\over {L_y}}$). This gives the strong impact of geometrical structures on the magnetic relaxation. To formalize the results, a power-law type of relation has been proposed between magnetic relaxation time of the system $\tau $ and the geometrical or structural factor, i.e., aspect ratio $(R)$ i.e, $\tau \sim R^{-s}$ and the relation is validated in the limit of large R. The thermal variation of the power exponent ($s$) can be fitted by a linear fuction i.e., $s(T ) = aT + b$. The aforementioned behaviors are universal in nature observed in both for the Metropolis and Glauber protocol as well as for both OBC and PBC. It is noteworthy to mention that such kind of power-law behaviour is observed in both for the cases of conserved area \cite{tikader2023effects} and non-conserved area of the lattice \cite{tikader2022effects}.  

For both cases of the deformations of the constant area and non-conserved area, the estimated values of $a$ and $b$ provided in Table-1 are almost insensible to the dynamical rules used here (Metropolis and Glauber dynamics). In contrary they exhibit strong dependence on the boundary conditions. The system with periodic bondary condition demands  comperatively higher magnitudes of $a$ and $b$. 

The relaxing nature of the magnetisation in 2D Ising ferromagnetic system (or model ferromagnetic monolayer) has been found to depend strongly on the dynamical rules also as well as the applied boundary conditions implemented in the MC simulational study.
The system with open boundary condition (OBC) yields relatively faster relaxation as compared to the PBC irrespective of the dynamical algorithms employed here. Metropolis algorithm also accelerates the relaxation process faster compared to Galuber protocol for any kind of applied boundary condition (open or periodic).

Furthermore, we have conducted an in-depth analysis of the dynamical evolution of spin-flip density to delve into the microscopic dynamics of relaxation. To expose the system, the OBC is applied here in order to investigate the behavior of the spin flip density in the core and surface part of 2D lattice. The evolution of spin-flip density is observed in the ordered phase or below $T_c$ temperature region and  disordered phase or above $T_c$ using different dynamical protocols. Interestingly, the saturated spin-flip density shows a logarithmic type of size dependence ($f_{sd} = a + b~ \log(L)$), for core part as well as surface part. These properties of saturated spin-flip density are observed in both ordered and disordered regimes. This is just to know whether the long range ferromagnetic order affects the spin-flip dynamics. The best fitted values of the parameters associated with logarithmic type of data fit and the statistical $\chi^2 $ are systematically provided in the Table-2 and Table-3 for the resulting data by Glauber protocol and Metropolis algorithm respectively. These findings are communicated in the paper by Tikader, Mallick and Acharyya, 2023.

This research-work servers as a compelling invitation to the experimentalists for systematic further explorations and analysis in the real-life usage of magnetic thin films or magnetic monolayers. It is highly probable that the \emph{magnetic coating} deposited on the non-magnetic substrate can exhibit analogous behaviour. In practical applications, the structural shape of \emph{magnetic coating} in magnetic stripe cards including credit cards, ATM cards, etc. could be varied to attain faster relaxation process for quick response. This suggests potential practical implications for enhancing the performance of magnetic systems in various real-world applications.

\vskip 1 cm

\section {Acknowledgements:} Ishita Tikader gratefully acknowledges the financial support from University Grant Commission, India.  Muktish Acharyya would like to acknowledges the FRPDF research Grant provided by Presidency University, Kolkata. The  fruitful collaboration with Olivia Mallick is gratefully acknowledged. We would like to thank Dr. Anuj Bhowmik for his kind help in preparing the manuscript.


\newpage

\bibliography{my_ref}
\bibliographystyle{apalike}

\end{document}